# Higher-order Topological Point State

Xiaoyin Li and Feng Liu*

Department of Materials Science and Engineering, University of Utah, Salt Lake City, Utah 84112, USA

Abstract

Higher-order topological insulators (HOTIs) have attracted increasing interest as a unique class of topological quantum materials. One distinct property of HOTIs is the crystalline symmetry-imposed topological state at the lower-dimensional *outer* boundary, e.g. the zero-dimensional (0D) corner state of a 2D HOTI, used exclusively as a universal signature to identify higher-order topology but yet with uncertainty. Strikingly, we discover the existence of *inner* topological point states (TPS) in a 2D HOTI, as the embedded "end" states of 1D first-order TI, as exemplified by those located at the vacancies in a Kekulé lattice. Significantly, we demonstrate that such inner TPS can be unambiguously distinguished from the trivial point-defect states, by their unique topology-endowed inter-TPS interaction and correlated magnetic response in spectroscopy measurements, overcoming an outstanding experimental challenge. Furthermore, based on first-principles calculations, we propose γ-graphyne as a promising material to observe the higher-order TPS. Our findings shed new light on our fundamental understanding of HOTIs, and also open an avenue to experimentally distinguishing and tuning TPS in the interior of a 2D sample for potential applications.

*Corresponding author.
fliu@eng.utah.edu

Topological insulators (TIs) featured with insulating bulk and topology-protected metallic boundary states have been extensively studied in the last two decades, because of their intriguing properties and promising potential for next-generation quantum technologies [1,2]. By the bulk-boundary correspondence principle, the ($n$-1)-dimensional [($n$-1)D] metallic boundary states of the first-order TIs are associated with the $n$D insulating bulk states around the topological gap, e.g. the 1D edge (2D surface) states in 2D (3D) TIs. The first-order topology is characterized with $Z_2$ invariant defined by parity inversion of bands. An interesting extension is the higher-order TI (HOTI) where in-gap topological states appear at the lower-than-($n$-1)D boundary, e.g. 0D corners of 2D HOTIs [3-9], which arise from crystalline symmetry as a generalization of topological crystalline insulator [10], characterized with topological invariant defined by eigenvalues of spatial symmetry operators. In other words, $n$D HOTIs have gapped ($n$-1)D boundary states but the band degeneracy is preserved at the ($n$-2)D or lower-dimensional boundary by spatial symmetry [7]. For example, in a 2D HOTI, corner states emerge at the intersection of two edges having mirror-symmetry-enforced opposite Dirac masses [11-14], and at all the corners subjected to $C_m$ rotation symmetry with a topological charge of $e/m$ [9,15].

The 0D corner state has been thought as a universal signature to identify 2D HOTIs [16-23]; however, its experimental confirmation remains elusive. To date, high-order topology in 2D has only been experimentally shown in artificial systems, e.g. photonic crystals or topological circuits [24-34]. One outstanding challenge is that a corner of a 2D atomic structure exhibits always a localized trivial electronic state, i.e., a dangling-bond state, which shows up in the spectroscopy measurements as a peak at the zero energy in the gap, independent of topology. This ambiguity has also caused controversy in identifying Majorana modes, sought-after for quantum computing [35], supposed to be located at the "ends" of a 1D topological superconducting nanowire [36,37].

In this Letter, we demonstrate the striking existence of *inner* topological point states (TPS), in contrast to *outer* corner states in a 2D HOTI, using the examples of TPS located at the vacancies in a Kekulé lattice. This is invoked by understanding the HOTI from a different angle, based on the concept of collective lattice coupling (CLC) [38] instead of symmetry-imposed topological invariant, in that a 2D HOTI is viewed as a 2D system embedded with 1D first-order TIs along selected directions [38]. Specifically in a Kekulé lattice, we identify those embedded 1D armchair directions to be topological nontrivial, following the 1D Su-Schriffer-Heeger (SSH) model [39] with a negative CLC [38]. Consequently, the vacancies sitting at the ends of the embedded 1D first-order topological chains host localized inner TPS in a 2D bulk sample.

Significantly, we show that such inner TPS can be unambiguously distinguished from the trivial point-defect states, by their unique topology-endowed inter-TPS interaction and correlated magnetic response in spectroscopy measurements, overcoming an outstanding experimental challenge. By varying the distance between two TPS, an energy gap opens when the distance is shorter than the twice of TPS penetration depth, a property absent for trivial defect states. Also, by applying a local magnetic probe as a modulator on one of the two interacting TPS, their long-range correlation can be directly detected by using another nonmagnetic probe on the other TPS, confirming irrefutably the high-order topology. Based on density functional theory (DFT) calculations, we further show that γ-graphyne [40-42], a 2D material with Kekulé lattice, is a promising platform to verify the proposed higher-order TPS and characterize their interaction and magnetic response. Below we will focus on the TPS realized in the Kekulé lattice/material, leaving the other TPS lattice models, including non-bipartite lattices, in Supplemental Material (SM) [43].

To illustrate the concept of higher-order TPS, we first adopt the spinless Kekulé lattice model shown in Fig. 1(a), whose tight-binding (TB) Hamiltonian is $H = -t_{\text{intra}} \sum_{\langle ij \rangle_{\text{intra}}} c_i^\dagger c_j - t_{\text{inter}} \sum_{\langle ij \rangle_{\text{inter}}} c_i^\dagger c_j$, with intra- and inter-cell ($t_{\text{intra}}$ and $t_{\text{inter}}$) nearest-neighbor hoppings. For $|t_{\text{intra}}| < |t_{\text{inter}}|$ ($|t_{\text{intra}}| > |t_{\text{inter}}|$), the system lies in a HOTI (normal insulator) phase. In the HOTI phase, in-gap topological corner states emerge at all six corners of a hexagon flake, subject to $C_6$ rotation symmetry [15,17,51,52]. Differently, the corner states appear only at the two 120° corners but not the two 60° corners of a rhombus flake (see SM [43]) [17,49-53], which is harder to understand because both types of corners are subject to mirror symmetry but only one has a Dirac mass. Instead, however, they can be clearly understood based on the CLC model [38] that the Kekulé lattice is effectively a 2D lattice embedded with 1D topological chains with a negative CLC, equivalent to 1D SSH chains with $|t_{\text{intra}}| < |t_{\text{inter}}|$. Specifically, such chains are along three armchair directions, ending at all six corners of a hexagon flake, but only two 120° corners of a rhombus flake whose two 60° corners are ends of a zigzag chain which is trivial having no corner state (see [38] and SM [43]). This understanding has then led us to realize that there can be *inner* TPS at vacancies, if they are at the ends of the embedded armchair chains, as demonstrated below.

To avoid the interference of *outer* corner states, we consider a sample of triangular flake of the Kekulé lattice with three 60° corners, as shown in Fig. 1(b), inside which two vacancies are created at the ends of an embedded armchair chain, separated by a distance L. Hypothetically, this can be thought as 1D SSH chain with two vacancies, as shown in Fig. 1(c), where the TPS emerge as end states. The middle section between the two vacancies is topological nontrivial with a negative CLC, having $t_{intra} < t_{inter}$; while the two outside sections on the left and right are topological trivial with a positive CLC, having $t_{intra} > t_{inter}$, because the vacancies effectively shift the position of two outside sections by half a unit cell relative to the middle section, as marked in Fig. 1(c). Effectively, the vacancies have cut the strong and weak bond at the ends of the middle and outside sections, respectively. TPS occur at the vacancies which are the 0D boundary between a topological (middle section) and trivial 1D bulk (two outside sections acting like vacuum).

Of course, practically, TPS cannot be realized in a 1D chain because the chain will be fragmented into pieces if atoms are removed to form vacancies. However, it can manifest as high-order TPS in a 2D structure with embedded 1D topological chains. Thus, to reveal its emergence, we calculate the eigen spectra of the triangular flakes by setting L = 14, $t_{intra}$ = 1 and $t_{inter}$ = 1.2 eV, respectively, as shown in Fig. 2(a), which are featured with two zero-energy states in the bulk gap. By analyzing the charge density distributions of these two states [inset of Fig. 2(a)], one sees that they are located around the vacancies, displaying a typical asymmetric shape of topological corner states (see Fig. S1(c) in SM [43]) [65], with a large penetration depth into the topological side of the armchair chain (the middle section). For comparison, we also calculate the energy spectrum for the same triangular flake in the normal insulator phase by setting $t_{intra}$ = 1.2 and $t_{inter}$ = 1 eV. The results are shown in Fig. 2(b), which are also featured with two zero-energy states inside the bulk gap; however, their charge densities are more symmetrically distributed around the vacancies and have a small penetration depth into the middle section [see the inset of Fig. 2(b)], indicating that they are trivial point-defect states.

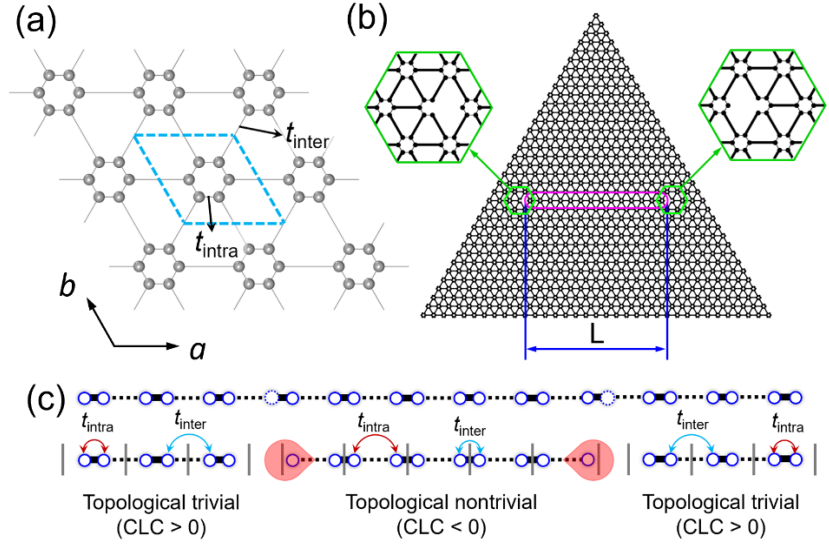

FIG. 1. (a) The Kekulé lattice with hopping $t_{intra}$ and $t_{inter}$. The unit cell is marked by the dashed rhombus. (b) Triangular flake with two vacancies separated by a distance L, which can be viewed as the ends of an embedded 1D armchair chain marked by the purple lines. The zoomed-in hexagons show geometrical details of vacancy. (c) Schematic 1D SSH model. Solid and dashed lines represent strong and weak bonds respectively. By hypothetically removing two atoms marked by the dashed circles in the top perfect chain, one creates TPS (red lobes) located at the vacancies of the bottom chain, which are effectively "ends" of the topological middle section with negative CLC, bounded by two trivial sections with positive CLC on the left and right.

The difference in charge density distribution around the defect, asymmetric vs. symmetric, is rooted in the fundamental difference between a TPS versus trivial point state in their decay mechanism and penetration length. The former corresponds to bulk Bloch wave function decaying exponentially with a large penetration length into the side of topological insulator bulk (the middle section), while the latter is more of a localized electronic state (i.e. dangling-bond state) having a small penetration depth. Specifically, the higher-order TPS is effectively the end state of the 1D first-order TI embedded in the 2D HOTI along specific crystalline directions. Thus, it must follow the bulk-boundary correspondence in 1D, namely the 0D TPS in real space has a small 1D dispersion within the bulk gap in momentum space. Consequently, it penetrates deeply into the nontrivial bulk side of the 1D armchair chain, but not so much into the trivial (vacuum) side. This makes not only an asymmetric distribution, but also a bulk-side penetration depth dependent on the size of the bulk gap that affects the dispersion, which can be tuned by hopping parameters. A smaller gap leads to a larger penetration depth and hence a longer-range inter-TPS interaction, as we show in Section V of SM [43]. This leads to longer-range interaction and correlation between two TPS at the ends of 1D topological bulk, which are topological features absent for trivial point-defect states.

Experimentally, the difference in charge density distribution around the defect, asymmetric vs. symmetric, might still be too subtle to distinguish. Therefore, we propose also to distinguish TPS from trivial point-defect states by varying their spatial separations and hence interactions. When two TPS have a distance smaller than the twice of their penetration length, they will interact with each other to open an energy gap, which can be directly observed by scanning tunneling spectroscopy (STS) [66] in the dI/dV curves as a function of separation. Whereas for trivial point-defect states, minimal energy splitting should be observed. In Fig. 2(c), we plot the evolution of two zero-energy states of Fig. 2(a) as a function of L. One clearly sees an energy splitting between the two TPS when L is reduced. In contrast, for the two zero-energy states of Fig. 2(b), the energy splitting is almost absent even at a very short L = 4 (Fig. 2(d)), since they are topologically trivial. We also calculate the energy spectra for flakes with different shapes and boundaries, and arbitrary vacancy locations within the flake, to confirm that the inner TPS are robust in all cases (Section VI of SM [43]).

We note that the proposed TPS, existing at specific sites protected by topology in any HOTI lattices including non-bipartite lattice, is fundamentally different from the trivial zero-energy state, existing at any missing site of a bipartite lattice that breaks the chiral symmetry. Besides 2D Kekulé lattice, TPS exist generally in HOTIs, including non-bipartite lattice models as we show for breathing Kagome [55] and puckered honeycomb lattice [23,56], and Kane-Mele model with an in-plane Zeeman field [57] (see SM [43]).

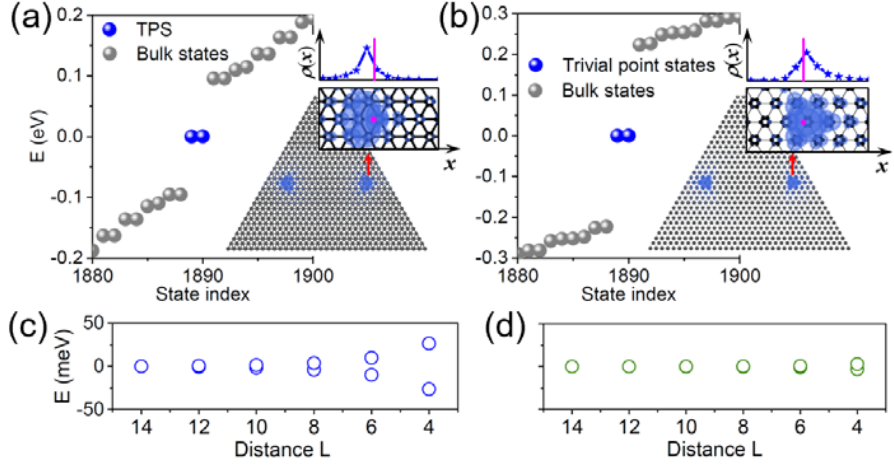

FIG. 2. Energy spectra for the triangular Kekulé flake with two (a) topological and (b) trivial point states. Insets are charge density distributions of in-gap point states highlighted in blue [$\rho(x)$ plots the line-cut along the 1D-chain direction], showing distinction between TPS and trivial point state. Purple lines and dots indicate the vacancy position. (c) The energy evolution of TPS as a function of spatial separation L. (d) Same as (c) but for trivial point states. For (a) and (b), L = 14.

Having established the distinguished presence of TPS in HOTI, we next explore the correlation effects of the TPS, which not only further characterizes their higher-order topology but also demonstrate their tunability with practical implications. In the above analyses, the distinction between Fig. 2(c) and (d) has to be made by measuring a series of samples with different vacancy separations, which can be experimentally challenging. Another viable approach is to use just one vacancy separation by detecting different responses to an external field between the TPS and trivial point-defect state via topology-endowed correlation effect. Since spin-orbit coupling (SOC) is negligible in most HOTI candidate materials, below we consider a spinful TB Hamiltonian of Kekulé lattice without SOC. In the top panel of Fig. 3(a), we show the DOS of TPS at L = 5 with other parameters set the same as in Fig. 2(a). There is a large energy splitting induced by the inter-TPS interaction. Also, there are two pairs of TPS for spin-up and -down electrons respectively, making the TPS to be tunable by applying a magnetic field. Specifically, by applying a local Zeeman field only on the left vacancy as shown in Fig. 1(b), the TPS on the right vacancy can be tuned, manifesting a quantum magnetic correlation between them. The local Zeeman field is modeled by adding a term $H_B = \sum_{i \in Z_B} c_i^\dagger \boldsymbol{B} \cdot \boldsymbol{s} c_i$ to the TB Hamiltonian, with $Z_B$ indicating the local region where the field is applied (see detail in Fig. S20 [43]). We use a Zeeman field along the z-direction with a strength of $\lambda_Z$, i.e., $\boldsymbol{B} = (0, 0, \lambda_Z)$, for illustration.

As shown in Fig. 3(a), by increasing the field strength (from top to bottom panel), one pair of TPS at the left vacancy are shifted away from the zero-energy, whereas the other pair of TPS at the right vacancy are modulated to approaching the zero-energy gradually. This process should be observable in dI/dV curves. The measured local DOS as a function of energy at the right vacancy is expected to evolve from exhibiting two peaks to one single peak near the Fermi energy with the increasing strength of local field, as shown from the top to bottom panel of Fig. 3(a). This can be understood as follows. For small L without field, the inter-TPS interaction leads to two DOS peaks bracketing the zero-energy for both TPS. When a local field is applied at one of the two defects, the TPS of that defect are tuned away from the zero-energy, so that its interaction with the other TPS is reduced and eventually vanishes for a strong enough field. Meantime, the TPS away from the field will regain their original zero-energy levels exhibiting one single DOS peak without interaction, as seen from the bottom panel of Fig. 3(a). This process can be described by an effective Hamiltonian, $H_{eff} = \varepsilon \tau_z + \gamma \tau_x + \frac{1}{2}\lambda_{eff}(1+\tau_z)\sigma_z$, where $\tau$ and $\sigma$ represent left/right TPS (L$_{TPS}$/R$_{TPS}$) and spin degree of freedom respectively. $\varepsilon$, $\gamma$ and $\lambda_{eff} = a\lambda_z$ are the onsite potential of TPS, inter-TPS interaction and effective Zeeman field strength, respectively, where $a$ is a constant depending on the areal coverage on the left TPS by the local field [67]. For the case of Fig. 3(a), $\varepsilon = 0$, $\gamma = 0.016$ and $a = 0.35$. The energy evolution of TPS as a function of $\lambda_z$ is calculated using $H_{eff}$, as shown in Fig. 3(b), which agree well with the direct solutions of TB Hamiltonian. This shows that the topology-endowed inter-TPS interaction renders them a correlated magnetic response, which can be used as a unique property to confirm their topological order and to tune their energy spectra. For comparison, we also performed similar calculations for the normal insulator phase with parameters set the same as those of Fig. 2(b), and the results are shown in Fig. 3(c). There is always one DOS peak at the zero-energy irrespective to the presence of a local Zeeman field, indicating that these point-defect states are topological trivial without correlated magnetic responses.

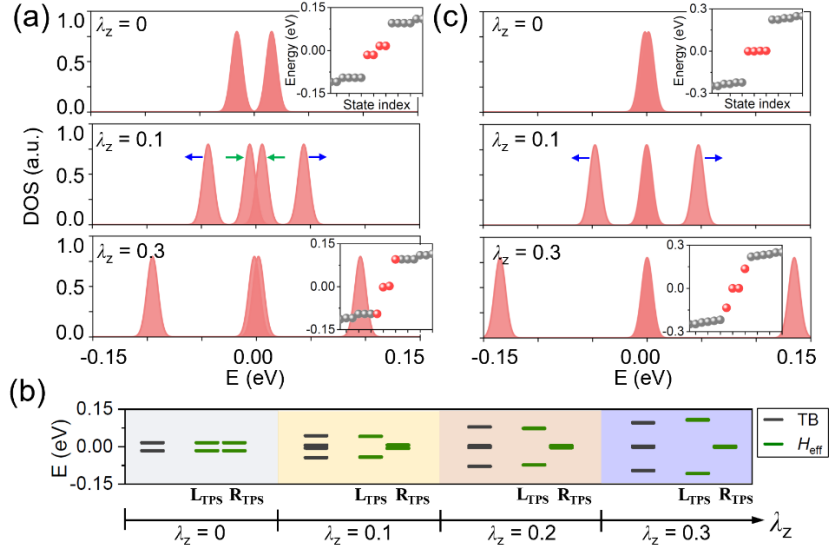

FIG. 3. (a) DOS of two TPS under a local Zeeman field of different strength for L = 5. In the middle panels, the blue (green) arrows indicate the trend of energy evolution of point states originated from the left (right) vacancy covered by (away from) the local field. Insets are energy spectra with point states highlighted by red. (b) Energy evolution of TPS as a function of $\lambda_Z$ calculated using TB and effective Hamiltonian, respectively. $L_{TPS}/R_{TPS}$ represents the TPS originated from the left/right vacancy. (c) Same as (a) but for trivial point states.

Lastly, we propose γ-graphyne as a promising material, based on DFT calculations (see Methods in SM [43]) to stimulate possible experiments. γ-graphyne was theoretically predicted a HOTI [17] as material realization of Kekulé lattice model; notably, its large-scale synthesis has been already achieved [42]. We construct a large supercell of γ-graphyne as shown in Fig. 4(a), and calculate its eigen-energy spectra at the Brillouin-zone-center Γ point, which are plotted in Fig. 4(b) and (c) for the pristine and defected case respectively. The point-defects (a vacancy cluster in this case) are configured following the end terminations of the first-order 1D TI of γ-graphyne [43]. Without defect, the spectrum simply inherits a bulk band gap of γ-graphyne; while with defects, there are two in-gap TPS [Fig. 4(c)] whose charge density are very asymmetrically, distributed only between the point-defects [Fig. 4(a)]. The energy evolution of TPS as a function of their spatial separation and the effect of a local Zeeman field are investigated by TB model analyses whose parameters are obtained by fitting the DFT results, both of which agree well with the above model studies (see Section IX of SM [43]). In addition to a local field as shown above, which can be difficulty to implement experimentally, one may also try to use a magnetic atom doping at one TPS, as we have illustrated using DFT calculations (see Fig. S24 in SM [43]).

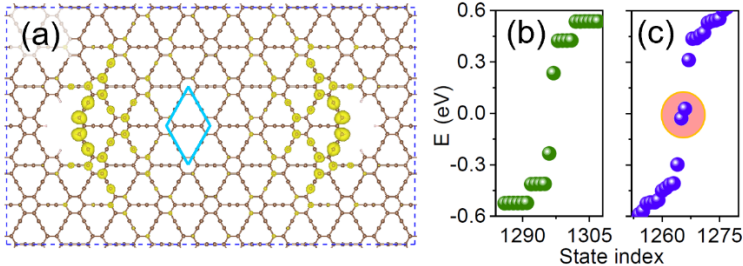

FIG. 4. (a) Supercell structure of γ-graphyne. The cyan rhombus marks the unit cell with lattice constant of $a = 6.89$ Å. (b) and (c) Energy spectra of γ-graphyne without and with point-defects, respectively. Charge density distributions of in-gap TPS [marked by the red ellipse in (c)] are plotted in (a) (yellow).

Before closing, we emphasize some advantages of the proposed TPS. Vacancies are the simplest and ubiquitously seen point defects in 2D materials [68-70]. Furthermore, vacancies may be created in a controllable manner, e.g. through atom manipulation by scanning tunneling microscopy [71] or adsorption of hydrogen or halogen atoms on selected atoms of a 2D sample (see Section X of SM [43]), to tune the number and location/separation of TPS (Fig. S26 [43]), which is much harder to do for outer corner states which would require changing the whole sample size and shape. This also enables TPS to occur in both topological and trivial systems with alternating bond strengths [72], further reducing the constraints on experimental synthesis. The TPS located in the interior of a 2D sample ease their detection and measurement, rather than working on the periphery of the sample, as well as suffer less influence from the environment. We also stress that the TPS we propose here is mechanistically different from the topological states bound to topological defects such as 0D disclinations and 1D dislocations, which act as fictitious flux centers and can induce $(n-2)$D topological state in both a $n$D first-order TI [73,74] and a $n$D HOTI [75-77]. Practically, generation of disclination and dislocation relies on formation of multiple grains with misaligned orientations or misfit strain, which is less controllable.

To close, our findings shed new light on our fundamental understanding of HOTIs with significant practical implications. The physical mechanism underlying the emergence of TPS is generally applicable, which can be extended to achieving 1D topological states bound to inner line defects instead of outer hinges of 3D HOTIs as well as inner Majorana point modes in a 2D HO topological superconductor (TSC) instead of those limited to corners of 2D TSC or ends of 1D TSC. The unambiguous distinction of *inner* TPS or Majorana point mode from a trivial point state and their high tunability pave the way to next-generation quantum technologies like quantum computing.

**Acknowledgement**

We thank H.Q. Huang, Z.F. Wang and Y.N. Zhou for helpful discussions. We acknowledge the support by U.S. DOE-BES (Grant No. DE-FG02-04ER46148). Computational resources of this work were supported by CHPC of University of Utah and DOE-NERSC.

**References**

[1] M. Z. Hasan and C. L. Kane, Rev. Mod. Phys. **82**, 3045 (2010).
[2] X.-L. Qi and S.-C. Zhang, Rev. Mod. Phys. **83**, 1057 (2011).
[3] W. A. Benalcazar, B. A. Bernevig, and T. L. Hughes, Science **357**, 61 (2017).
[4] Y. Peng, Y. Bao, and F. von Oppen, Phys. Rev. B **95**, 235143 (2017).
[5] W. A. Benalcazar, B. A. Bernevig, and T. L. Hughes, Phys. Rev. B **96**, 245115 (2017).
[6] Z. Song, Z. Fang, and C. Fang, Phys. Rev. Lett. **119**, 246402 (2017).
[7] F. Schindler, A. M. Cook, M. G. Vergniory, Z. Wang, S. S. P. Parkin, B. A. Bernevig, and T. Neupert, Sci. Adv. **4**, eaat0346 (2018).
[8] H. Huang, J. Fan, D. Li, and F. Liu, Nano Lett. **21**, 7056 (2021).
[9] C. Wang, F. Liu, and H. Huang, Phys. Rev. Lett. **129**, 056403 (2022).
[10] L. Fu, Phys. Rev. Lett. **106**, 106802 (2011).
[11] R. Jackiw and C. Rebbi, Phys. Rev. D **13**, 3398 (1976).
[12] M. Geier, L. Trifunovic, M. Hoskam, and P. W. Brouwer, Phys. Rev. B **97**, 205135 (2018).
[13] L. Trifunovic and P. W. Brouwer, Phys. Rev. X **9**, 011012 (2019).
[14] L. Trifunovic and P. W. Brouwer, Phys. Status Solidi B **258**, 2000090 (2021).
[15] W. A. Benalcazar, T. Li, and T. L. Hughes, Phys. Rev. B **99**, 245151 (2019).
[16] X.-L. Sheng, C. Chen, H. Liu, Z. Chen, Z.-M. Yu, Y. X. Zhao, and S. A. Yang, Phys. Rev. Lett. **123**, 256402 (2019).
[17] B. Liu, G. Zhao, Z. Liu, and Z. F. Wang, Nano Lett. **19**, 6492 (2019).
[18] E. Lee, R. Kim, J. Ahn, and B.-J. Yang, npj Quantum Mater. **5**, 1 (2020).
[19] M. J. Park, Y. Kim, G. Y. Cho, and S. Lee, Phys. Rev. Lett. **123**, 216803 (2019).
[20] H. Huang and F. Liu, Natl. Sci. Rev. **9**, nwab170 (2022).
[21] X. Ni, H. Huang, and J.-L. Brédas, J. Am. Chem. Soc. **144**, 22778 (2022).
[22] T. Hu, T. Zhang, H. Mu, and Z. Wang, J. Phys. Chem. Lett. **13**, 10905 (2022).
[23] S. Qian, C.-C. Liu, and Y. Yao, Phys. Rev. B **104**, 245427 (2021).
[24] M. Serra-Garcia, V. Peri, R. Süsstrunk, O. R. Bilal, T. Larsen, L. G. Villanueva, and S. D. Huber, Nature **555**, 342 (2018).
[25] X. Ni, M. Weiner, A. Alù, and A. B. Khanikaev, Nat. Mater. **18**, 113 (2019).
[26] H. Xue, Y. Yang, F. Gao, Y. Chong, and B. Zhang, Nat. Mater. **18**, 108 (2019).

[27] X. Zhang, H.-X. Wang, Z.-K. Lin, Y. Tian, B. Xie, M.-H. Lu, Y.-F. Chen, and J.-H. Jiang, Nat. Phys. **15**, 582 (2019).
[28] J. Noh, W. A. Benalcazar, S. Huang, M. J. Collins, K. P. Chen, T. L. Hughes, and M. C. Rechtsman, Nat. Photon. **12**, 408 (2018).
[29] S. Mittal, V. V. Orre, G. Zhu, M. A. Gorlach, A. Poddubny, and M. Hafezi, Nat. Photon. **13**, 692 (2019).
[30] A. El Hassan, F. K. Kunst, A. Moritz, G. Andler, E. J. Bergholtz, and M. Bourennane, Nat. Photon. **13**, 697 (2019).
[31] C. W. Peterson, W. A. Benalcazar, T. L. Hughes, and G. Bahl, Nature **555**, 346 (2018).
[32] S. Imhof, C. Berger, F. Bayer, J. Brehm, L. W. Molenkamp, T. Kiessling, F. Schindler, C. H. Lee, M. Greiter, T. Neupert, and R. Thomale, Nat. Phys. **14**, 925 (2018).
[33] W. Zhang, D. Zou, Q. Pei, W. He, J. Bao, H. Sun, and X. Zhang, Phys. Rev. Lett. **126**, 146802 (2021).
[34] S. N. Kempkes, M. R. Slot, J. J. van den Broeke, P. Capiod, W. A. Benalcazar, D. Vanmaekelbergh, D. Bercioux, I. Swart, and C. Morais Smith, Nat. Mater. **18**, 1292 (2019).
[35] C. Nayak, S. H. Simon, A. Stern, M. Freedman, and S. Das Sarma, Rev. Mod. Phys. **80**, 1083 (2008).
[36] R. M. Lutchyn, J. D. Sau, and S. Das Sarma, Phys. Rev. Lett. **105**, 077001 (2010).
[37] H. Zhang, C.-X. Liu, S. Gazibegovic, D. Xu, J. A. Logan, G. Wang, N. van Loo, J. D. S. Bommer, M. W. A. de Moor, D. Car, R. L. M. Op het Veld, P. J. van Veldhoven, S. Koelling, M. A. Verheijen, M. Pendharkar, D. J. Pennachio, B. Shojaei, J. S. Lee, C. J. Palmstrøm, E. P. A. M. Bakkers, S. D. Sarma, and L. P. Kouwenhoven, Nature **556**, 74 (2018).
[38] For details about the concept of collective lattice coupling (CLC) and its application in the 2D HOTI please see our previous paper [B. Xia, H. Liu, and F. Liu, arXiv preprint arXiv:2208.11764 (2022)].
[39] W. P. Su, J. R. Schrieffer, and A. J. Heeger, Phys. Rev. Lett. **42**, 1698 (1979).
[40] Y. Li, L. Xu, H. Liu, and Y. Li, Chem. Soc. Rev. **43**, 2572 (2014).
[41] Q. Li, Y. Li, Y. Chen, L. Wu, C. Yang, and X. Cui, Carbon **136**, 248 (2018).
[42] Y. Hu, C. Wu, Q. Pan, Y. Jin, R. Lyu, V. Martinez, S. Huang, J. Wu, L. J. Wayment, N. A. Clark, M. B. Raschke, Y. Zhao, and W. Zhang, Nat. Synth. **1**, 449 (2022).
[43] See Supplemental Material for details, which includes Refs [17,23,44-64].
[44] G. Kresse and J. Furthmüller, Phys. Rev. B **54**, 11169 (1996).
[45] P. E. Blöchl, Phys. Rev. B **50**, 17953 (1994).
[46] J. P. Perdew, K. Burke, and M. Ernzerhof, Phys. Rev. Lett. **77**, 3865 (1996).
[47] J. P. Perdew, K. Burke, and M. Ernzerhof, Phys. Rev. Lett. **78**, 1396 (1997).
[48] H. J. Monkhorst and J. D. Pack, Phys. Rev. B **13**, 5188 (1976).
[49] L.-H. Wu and X. Hu, Sci. Rep. **6**, 24347 (2016).
[50] Y. Liu, C.-S. Lian, Y. Li, Y. Xu, and W. Duan, Phys. Rev. Lett. **119**, 255901 (2017).
[51] F. Zangeneh-Nejad and R. Fleury, Phys. Rev. Lett. **123**, 053902 (2019).
[52] T. Mizoguchi, H. Araki, and Y. Hatsugai, J. Phys. Soc. Jpn. **88**, 104703 (2019).
[53] Y. Zhou and R. Wu, Phys. Rev. B **107**, 035412 (2023).
[54] S.-Q. Shen, *Topological insulators* (Springer, 2012), Vol. 174.
[55] M. Ezawa, Phys. Rev. Lett. **120**, 026801 (2018).
[56] M. Pan, D. Li, J. Fan, and H. Huang, Npj Comput. Mater. **8**, 1 (2022).
[57] Y. Ren, Z. Qiao, and Q. Niu, Phys. Rev. Lett. **124**, 166804 (2020).
[58] W. A. Harrison, *Electronic Structure and the Properties of Solids: The Physics of the Chemical Bond* (Dover Publications, Newburyport, 2012).
[59] C. L. Kane and E. J. Mele, Phys. Rev. Lett. **95**, 146802 (2005).
[60] C. L. Kane and E. J. Mele, Phys. Rev. Lett. **95**, 226801 (2005).
[61] R. Wiesendanger, D. Bürgler, G. Tarrach, T. Schaub, U. Hartmann, H. J. Güntherodt, I. V. Shvets, and J. M. D. Coey, Appl. Phys. A **53**, 349 (1991).
[62] Z. Zhang, X. Ni, H. Huang, L. Hu, and F. Liu, Phys. Rev. B **99**, 115441 (2019).
[63] M. Zhou, W. Ming, Z. Liu, Z. Wang, P. Li, and F. Liu, Proc. Natl. Acad. Sci. U.S.A. **111**, 14378 (2014).
[64] H. Zhang, Y. Wang, W. Yang, J. Zhang, X. Xu, and F. Liu, Nano Lett. **21**, 5828 (2021).

[65] N. Arai and S. Murakami, J. Phys. Soc. Jpn. **90**, 074711 (2021).
[66] J. A. Stroscio, R. M. Feenstra, and A. P. Fein, Phys. Rev. Lett. **57**, 2579 (1986).
[67] This is because the applied external Zeeman field only covers a local spatial region of the sample and a partial wavefuntion of the left TPS, which makes $a < 1$.
[68] F. Banhart, J. Kotakoski, and A. V. Krasheninnikov, ACS Nano **5**, 26 (2011).
[69] Z. Lin, B. R. Carvalho, E. Kahn, R. Lv, R. Rao, H. Terrones, M. A. Pimenta, and M. Terrones, 2D Mater. **3**, 022002 (2016).
[70] J. Hong, C. Jin, J. Yuan, and Z. Zhang, Adv. Mater. **29**, 1606434 (2017).
[71] D. Wong, J. Velasco, L. Ju, J. Lee, S. Kahn, H.-Z. Tsai, C. Germany, T. Taniguchi, K. Watanabe, A. Zettl, F. Wang, and M. F. Crommie, Nat. Nanotechnol. **10**, 949 (2015).
[72] For systems with alternating bond strengths like the 1D SSH and 2D Kekulé lattices, because the position of vacancy is not constrained, one can shift the relative position of the created vacancies to make the middle section between the two vacancies topological nontrivial, for both the topological nontrivial and trivial phases of the bulk. Therefore, the proposed TPS can emerge in both bulk phases, as demonstrated in Section XI of Supplemental Material.
[73] Y. Ran, Y. Zhang, and A. Vishwanath, Nat. Phys. **5**, 298 (2009).
[74] A. Rüegg and C. Lin, Phys. Rev. Lett. **110**, 046401 (2013).
[75] T. Li, P. Zhu, W. A. Benalcazar, and T. L. Hughes, Phys. Rev. B **101**, 115115 (2020).
[76] Y. Liu, S. Leung, F.-F. Li, Z.-K. Lin, X. Tao, Y. Poo, and J.-H. Jiang, Nature **589**, 381 (2021).
[77] C. W. Peterson, T. Li, W. Jiang, T. L. Hughes, and G. Bahl, Nature **589**, 376 (2021).

Supplemental Material for

# Higher-order Topological Point State

Xiaoyin Li and Feng Liu*

Department of Materials Science and Engineering, University of Utah, Salt Lake City, Utah 84112, USA

**Table of Contents**

## I. Methods for first-principles calculations

Density functional theory (DFT) calculations are performed by using Vienna *Ab initio* Simulation Package [1] to investigate the structural and electronic properties of γ-graphyne. The projector augmented wave method [2] is used to describe the interactions between core-valence electrons. The Perdew-Burke-Ernzerhof (PBE) functional within the generalized gradient approximation (GGA) [3,4] is used to treat the exchange-correlation interactions and the energy cutoff is set to 520 eV. The Brillouin zone is sampled by a *k*-point mesh of $7 \times 7 \times 1$ ($1 \times 1 \times 1$) for unit cell (supercell) following the Monkhorst-Pack scheme [5]. The convergence criterion for energy in the self-consistent field iterations is set to $10^{-5}$ eV. To account for the strong onsite Coulomb repulsion of the *d* shell electrons of V, the DFT+U method is implemented for the calculation of V-adsorbed γ-graphyne with the effective U value being 2 eV.

## II. Higher-order topological insulator realized in Kekulé lattice model

It has been previously demonstrated that the Kekulé lattice model [see Fig. 1(a) of the main text] with $|t_{intra}| < |t_{inter}|$ is a higher-order topological insulator (HOTI), and with $|t_{intra}| > |t_{inter}|$ is a normal insulator [6-11]. By setting $t_{intra} = 1$ and $t_{inter} = 1.2$ eV, we calculate the bulk energy band structure and energy spectrum of a rhombus flake. The results in Fig. S1(a) and (b) show that, in the energy spectrum of the flake, there are two states in the bulk band gap. Analyses of charge density distributions verify that these two states are topological corner states located at the 120° corners of the rhombus flake, confirming a HOTI. We also calculate the corresponding bulk energy band structure, energy spectrum of the rhombus flake and charge density distributions of states near the band gap by setting $t_{intra} = 1.2$ and $t_{inter} = 1$ eV. The results are plotted in Fig. S1(d)-(f). One can see that, for such electron hoppings, the system is in the normal insulator phase. These results are all consistent with previous studies.

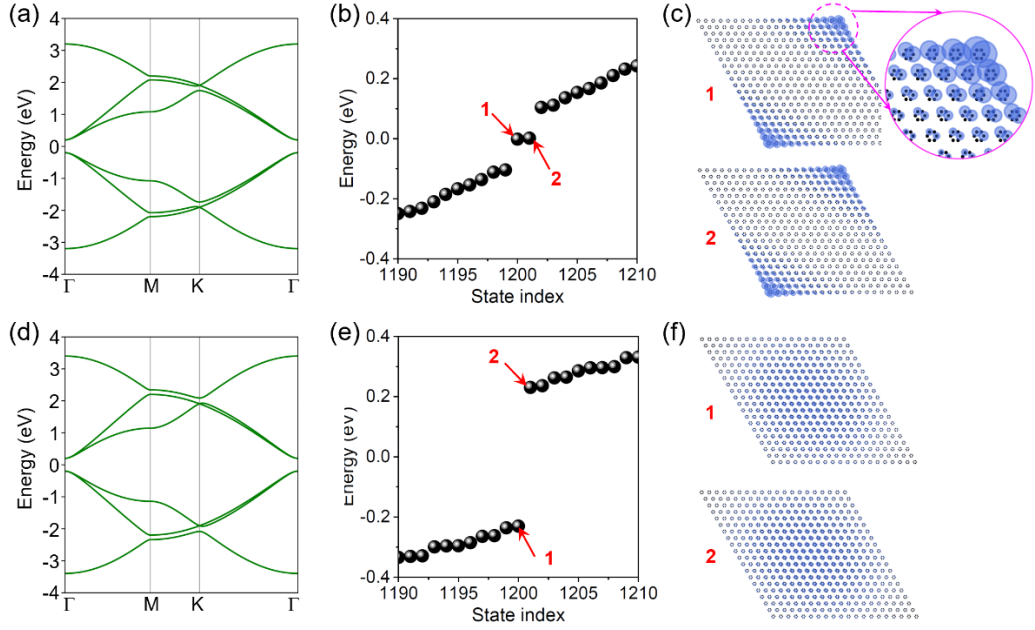

FIG. S1. (a)-(c) are the bulk electronic band structure, energy spectrum of the finite rhombus flake, and charge density distributions of state 1 and state 2 as indicated in (b) of the Kekulé lattice, respectively. Here the lattice is in the HOTI phase with $t_{intra} = 1$ and $t_{inter} = 1.2$ eV. (d)-(f) are same as (a)-(c) but for a normal insulator phase with $t_{intra} = 1.2$ and $t_{inter} = 1$ eV.

## III. Topological end states of nanoribbon and topological corner states of nanoflake

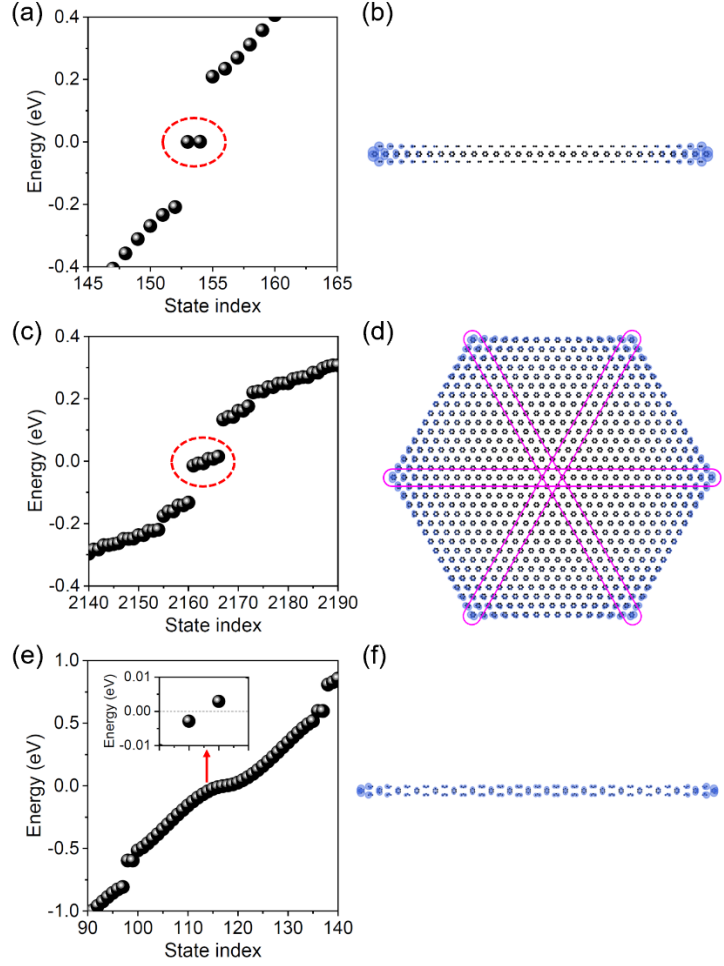

FIG. S2. Illustration of the first-order topologically nontrivial nanoribbon and its relationship with topological corner states of the 2D finite flake by using the Kekulé lattice as an example. Here the Kekulé lattice is in the HOTI phase with $t_{intra} = 1$ and $t_{inter} = 1.2$ eV. (a) Energy spectrum of a finite 1D armchair nanoribbon as embedded in a 2D Kekulé lattice. It effectively realizes the 1D Su-Schriffer-Heeger (SSH) model and exhibits two topological end states as highlighted by the red dashed ellipse. The spatial distribution of the topological end states is shown in (b). (c) and (d) are energy spectra of a hexagonal flake and the spatial distribution of its topological corner states, respectively, which are highlighted by the red dashed ellipse in (c). One can see that these corner states can be viewed as the combination of the topological end state subjected to $C_6$ rotation symmetry. (e) Energy spectrum of a finite 1D zigzag nanoribbon, which is metallic and topological trivial. Inset of (e) is the zoomed-in spectrum near the zero-energy, and the spatial distribution of the two states are plotted in (f).

## IV. Determining the first-order topologically nontrivial nanostructure

A convenient way to determine the first-order topologically nontrivial 1D nanostructure embedded in the bulk of 2D HOTI is finding a pair of topological corner states of a finite flake of the 2D HOTI. Then, the line connecting the paired topological corner states will indicate the periodic direction of the first-order topologically nontrivial 1D nanostructure, and one should obtain topological end states if the nanostructure is cut from the 2D flake into a finite nanoribbon. Here, the illustrations of the Kekulé lattice model are shown in Fig. S3 as an example. Demonstrations for other lattice models can be found in the following Section VII.

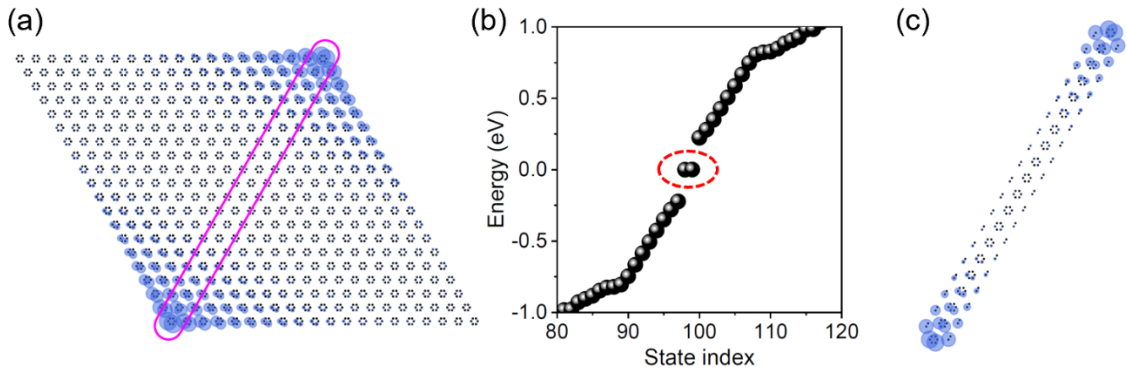

FIG. S3. Determination of the first-order topologically nontrivial 1D nanostructure of the 2D HOTI in the Kekulé lattice model. (a) The spatial distribution of a pair of topological corner states for a rhombus flake of the Kekulé lattice. The line connecting the paired topological corner states indicates the periodic direction of the first-order topologically nontrivial 1D nanostructure, as marked by the purple line. (b) Energy spectrum for the finite 1D nanoribbon which is cut from the 2D flake of (a). The nanoribbon realizes the first-order topologically nontrivial 1D SSH model and exhibits two topological end states highlighted by the red dashed line, whose charge density distributions are displayed in (c). Here the Kekulé lattice is in the HOTI phase with $t_{intra} = 1$ and $t_{inter} = 1.2$ eV.

## V. Penetration depth of topological point state under different $t_{inter}/t_{intra}$ ratios

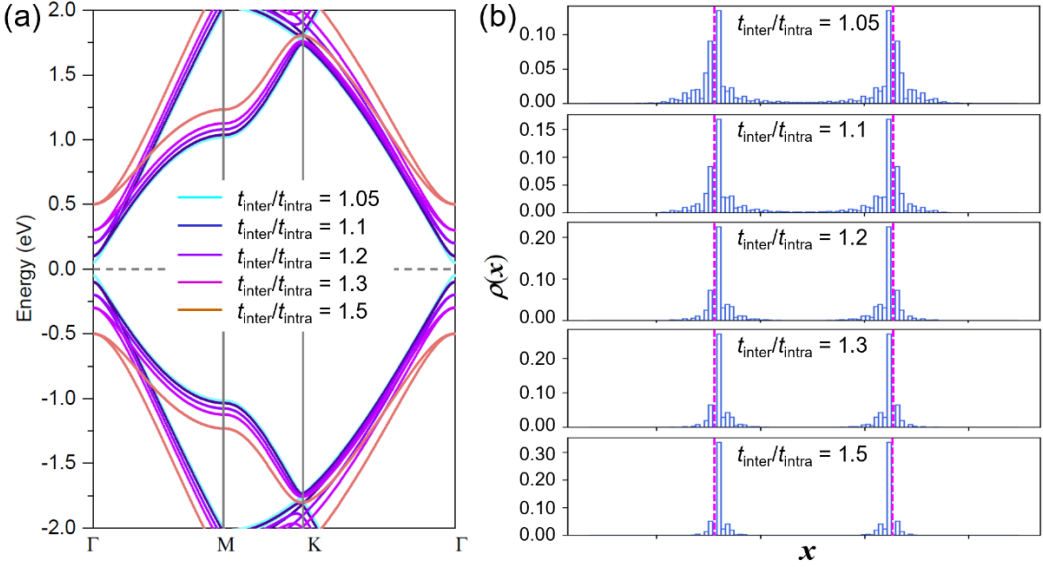

FIG. S4. (a) Bulk electronic band structure of Kekulé lattice with different $t_{inter}/t_{intra}$ ratios. For the convenience of comparison, a fixed value of 1 eV is set for $t_{intra}$. (b) The corresponding charge density distribution functions [$\rho(x)$] of TPS. The finite flake used for the calculation of TPS is the same as that in Fig. 2(a) of the main text and $x$ is along the 1D-chain direction inside the flake. One can see that a smaller $t_{inter}/t_{intra}$ ratio leads to a smaller topological band gap and a larger penetration depth of the generated TPS. These results are consistent with topological 1D SSH chain studies, where a smaller $t_{inter}/t_{intra}$ ratio will lead to a larger penetration depth of the topological end state [12]. We want to stress that the larger penetration depth of TPS can only give rise to a finite inter-TPS interaction over a longer-range, while the resulted energy splitting between the two TPS also depends on the amplitude of their wavefunction overlapping. In principle, all these properties of real materials can be accurately predicted by first-principle calculations, as a reliable reference for experimental observations.

## VI. Robustness of topological point states

To check the robustness of topological point states (TPS), we first construct the point-defects in flakes with different shapes and calculate corresponding energy spectra with results summarized in Fig. S5-S8. One can see that the TPS are preserved for all the cases, demonstrating the robustness of inner TPS against changing outer boundary shapes, which will ease their experimental synthesis and characterization.

We also consider the point-defects nonsymmetrically distributed relative to the finite triangular sample [see Fig. S9(a)] and missing atoms in a nonsymmetric manner [see Fig. S9(d)].

The results of two well-separated point-defects are summarized in Fig. S9. One can see that the TPS are preserved in both cases, demonstrating the robustness of TPS as long as the chain connecting the two TPS remains intact as a first-order topological insulator. We also check the results for two point-defects with reduced spatial separation as shown in Fig. S10 and demonstrate that the longer-range communications between the two TPS are well maintained in both cases.

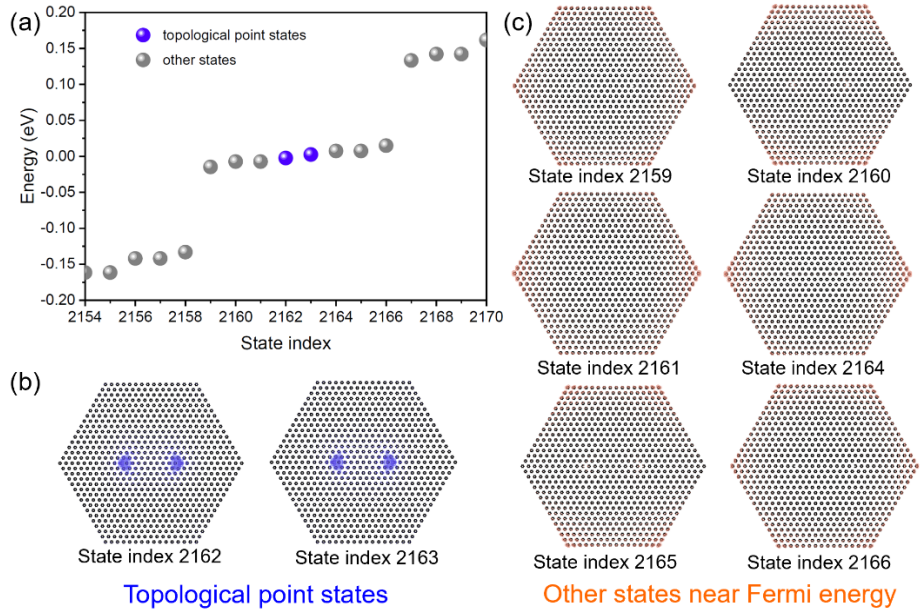

FIG. S5. (a) Energy spectrum of a hexagonal flake with TPS. (b) and (c) are charge density distributions of the TPS and the other states near the Fermi energy, represented by blue and orange respectively. The system is in the HOTI phase with $t_{intra} = 1$ and $t_{inter} = 1.2$ eV.

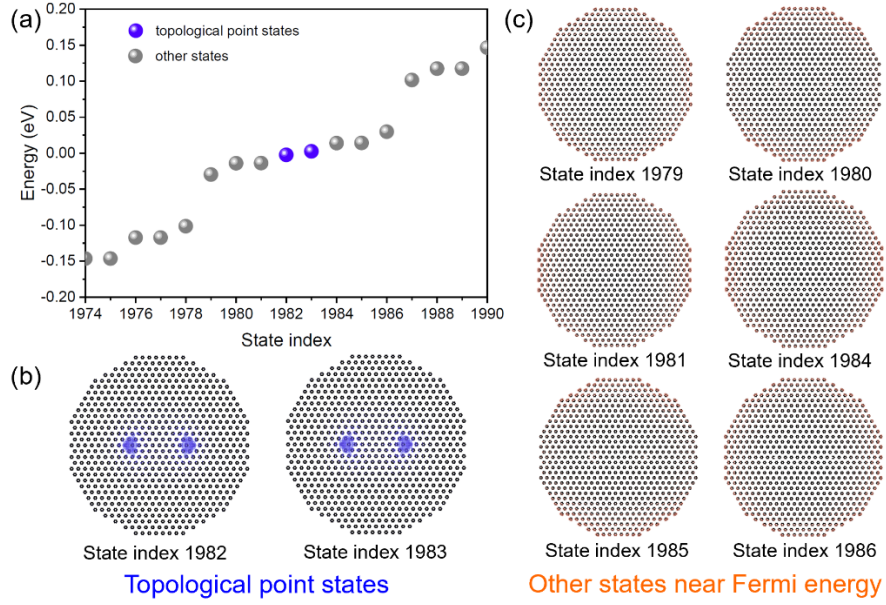

FIG. S6. Same as Fig. S5 but for a flake with a polygonal shape.

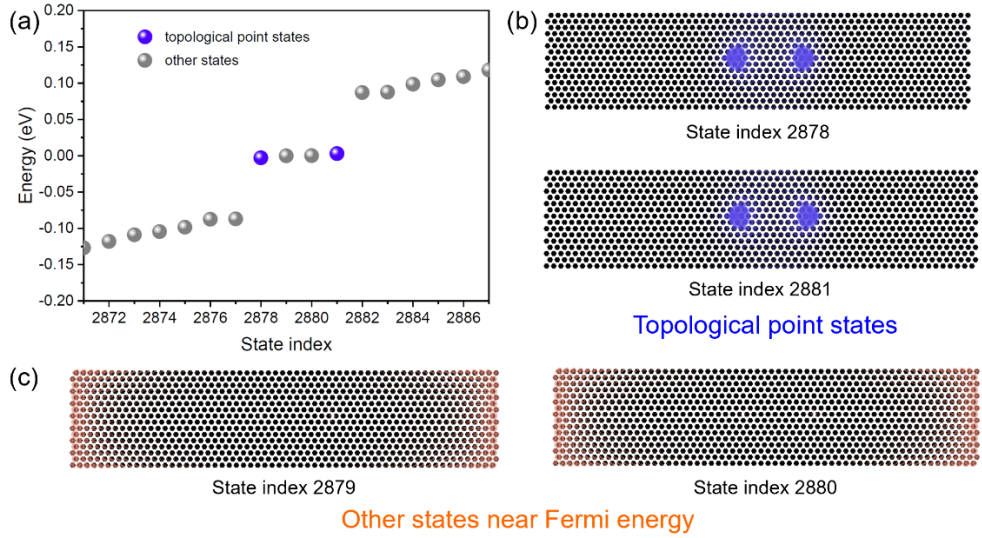

FIG. S7. Same as Fig. S5 but for a flake with a rectangular shape.

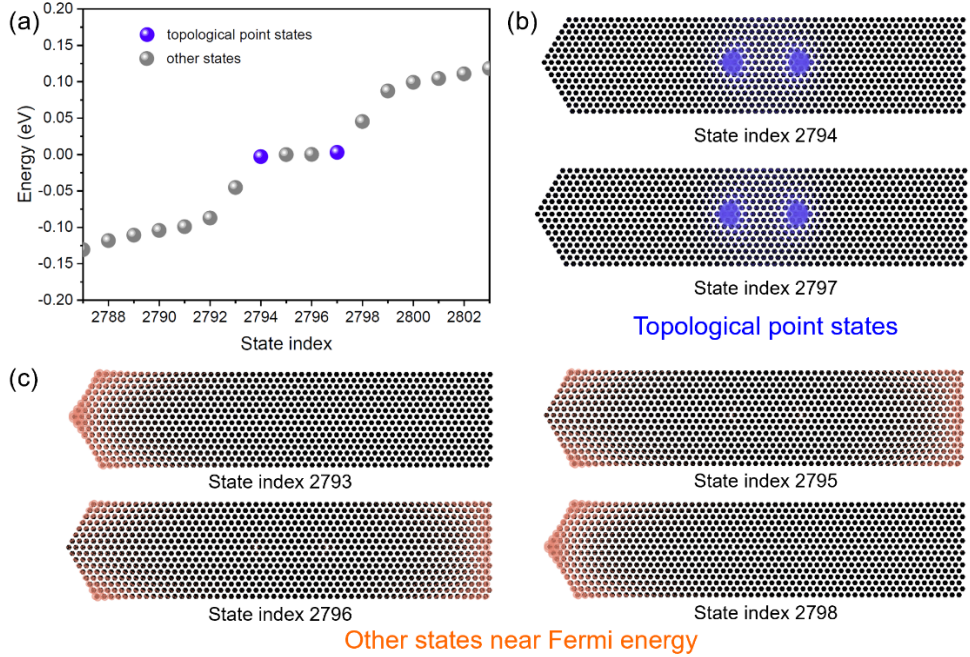

FIG. S8. Same as Fig. S5 but for a flake with irregular shape.

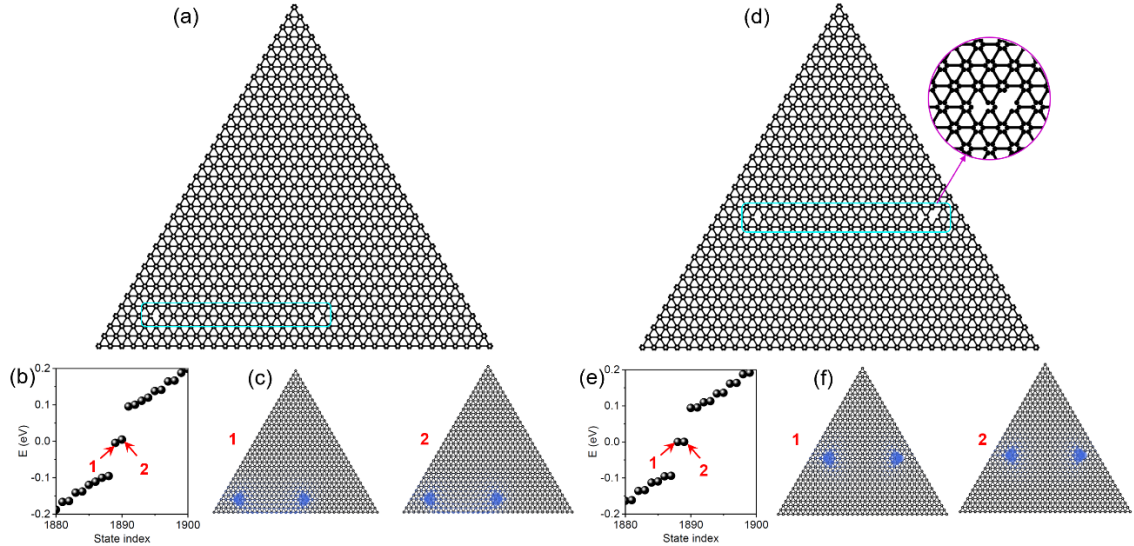

FIG. S9. (a) The finite flake with two point-defects nonsymmetrically distributed relative to the whole sample. The point-defects are highlighted by the cyan rectangle. (b) Energy spectrum of the finite flake as shown in (a), exhibiting in-gap TPS marked as 1 and 2. (c) Charge density distributions of in-gap TPS 1 and 2 (blue). (d)-(f) Same as (a)-(c) but for the finite flake with two point-defects and missing atoms in a nonsymmetric manner. The inset shows structural detail near the right point-defect. For both cases, the bulk Kekulé lattice is in the topological nontrivial phase with $t_{intra} = 1$ and $t_{inter} = 1.2$ eV, and the spatial separation between two point-defects is L = 14.

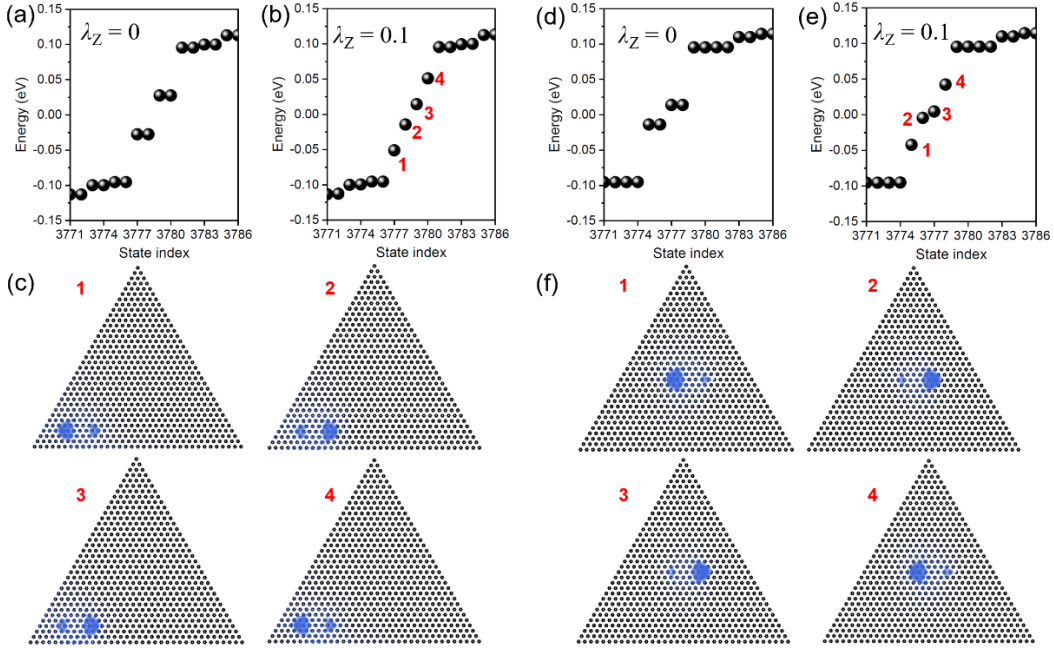

FIG. S10. (a) Energy spectrum of the finite flake with two point-defects whose configurations are the same as those of Fig. S9(a) but with a smaller spatial separation (L = 5). In this situation, TPS from two point-defects can interact with each other and exhibit an energy splitting as shown in the energy spectrum. (b) Same as (a) but with a local Zeeman field ($\lambda_Z = 0.1$ eV) applied on the left point-defect. (c) Charge density distributions of the four TPS marked as 1, 2, 3 and 4 (blue). One can see that the applied local Zeeman field tunes the two left TPS away from the Fermi energy, and consequently the two right TPS tend to regain their original zero-energy levels and the energy splitting between them becomes smaller. (d)-(f) Same as (a)-(c) but for the point-defects whose configurations are the same as those of Fig. S9(d). For all cases, the bulk Kekulé lattice is in the topological nontrivial phase with $t_{intra} = 1$ and $t_{inter} = 1.2$ eV, and the spin degree of freedom is included.

## VII. Topological point states realized in other lattice models

To confirm the generality of the proposed higher-order TPS, we explore their realization in some other 2D lattice models. Here we show TPS in three representative lattice models, namely, the breathing Kagome lattice [13], the puckered honeycomb lattice [14,15] and the Kane-Mele model with an in-plane Zeeman field [16].

Fig. S11(a) shows the structure of the breathing Kagome lattice with nearest-neighbor (NN) hoppings $t_{intra}$ and $t_{inter}$. For $-1 < t_{intra}/t_{inter} < ½$, the system is in the HOTI phase and its triangular flake exhibits three zero-energy states located at the 60° corners [13]. Here we set $t_{intra} = 0.3$ and $t_{inter} = 1$ eV to realize the HOTI phase. Fig. S11(b) and (c) are the calculated bulk energy band structure and energy spectrum of the finite triangular flake. One can see that there are three zero-energy states emerged for the triangular flake. Analyzing charge density distributions of these states (Fig. S11(d)) confirms that they are topological corner states located at the 60° corners of the flake.

By connecting two 60° corners, one is supposed to obtain the topologically nontrivial 1D nanostructure of the breathing Kagome lattice. In Fig. S12(a), we plot the energy spectrum of a finite nanoribbon cut from the 2D triangular flake, which shows the emergence of two states inside the band gap. Analysis of the spatial distribution of these two states demonstrates that they are topological end states localized at the ends of the finite nanoribbon, as shown in Fig. S12(b). To generate TPS in the breathing Kagome lattice, we construct the point-defects by following the end terminations of the finite nanoribbon and obtain the supercell structure containing two point-defects as shown in Fig. S12(c). Here periodic boundary conditions are adopted along the *a* and *b* directions to avoid the distraction from topological corner states. By solving the tight-binding (TB) Hamiltonian of the supercell structure, we obtain energy spectrum in Fig. S12(d), where two degenerate states emerge inside the bulk band gap. Charge density distributions of these two states (Fig. S12(e)) demonstrate that they are TPS located at the point-defects. Based on our analyses in the main text, we know that the topological corner states of finite flakes are effectively the topological end states of first-order topologically nontrivial 1D nanostructures embedded in the bulk of 2D HOTI, which can facilitate the creation of TPS by just following the corner terminations of corresponding flakes. In Fig. S13(a), we show three point-defects inside the supercell structure, which are constructed by following the 60° corner terminations of the triangular flake. The corresponding energy spectrum is present in Fig. S13(b), where three zero-energy states emerge inside the bulk band gap. Charge density distributions of these three states demonstrate that they are also TPS located at the point-defects, as shown in Fig. S13(c).

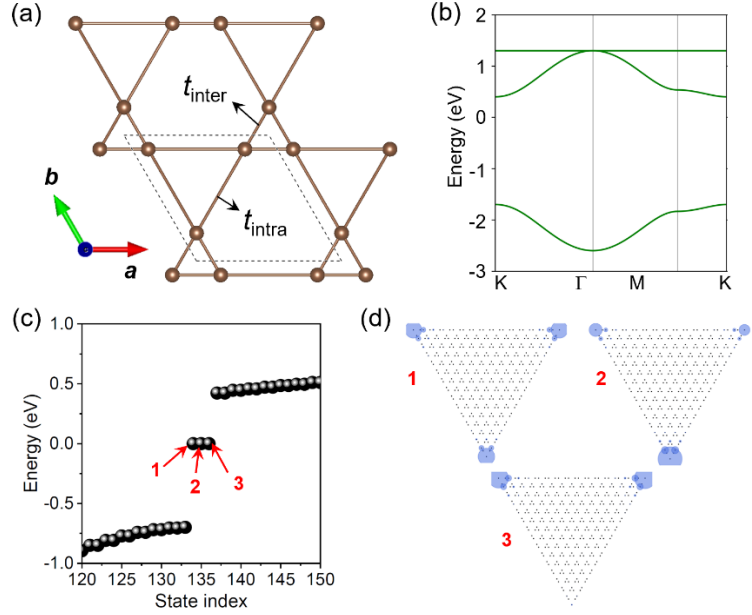

FIG. S11. (a) Structure of the breathing Kagome lattice with hoppings of $t_{intra} = 0.3$ and $t_{inter} = 1$ eV. The unit cell is represented by the dashed rhombus. (b) Bulk energy band structure of the breathing Kagome lattice. (c) Energy spectrum of the finite triangular flake of the breathing Kagome lattice, exhibiting three in-gap topological corner states. (d) Charge density distributions of topological corner states 1, 2, 3 as indicated in (c).

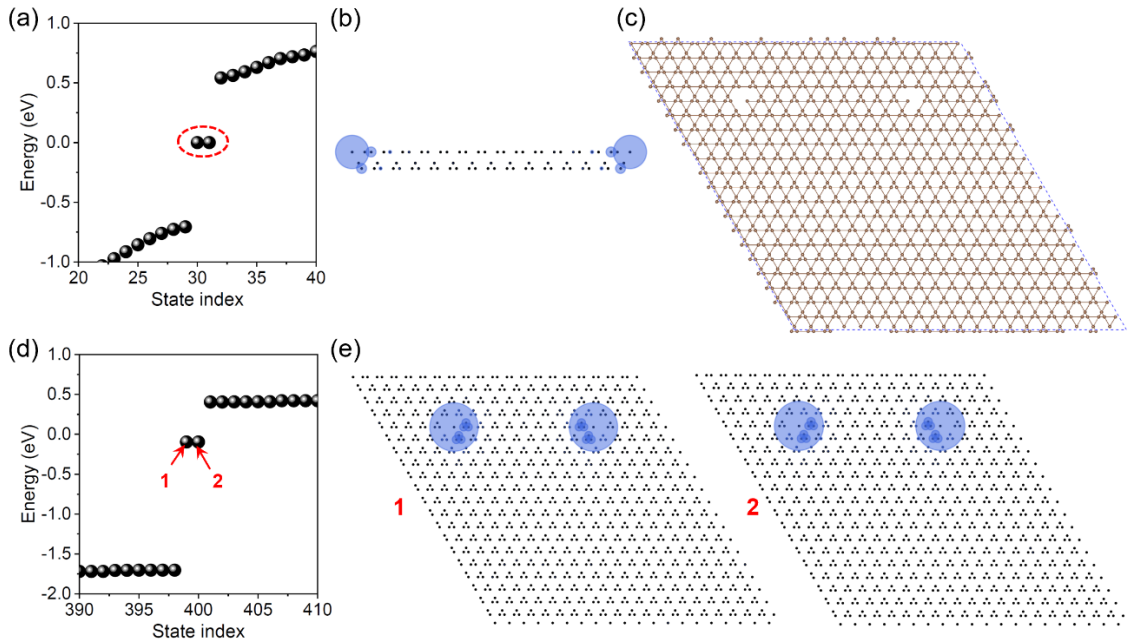

FIG. S12. (a) Energy spectrum of the finite 1D nanoribbon of the breathing Kagome lattice, showing the emergence of two topological end states inside the bulk band gap. (b) Charge density distribution of the two topological end states. (c) Supercell structure of the breathing Kagome

lattice containing two point-defects, which are constructed to be effectively the ends of the finite 1D nanoribbon. Here the periodic boundary conditions are adopted along the *a* and *b* directions to avoid the distraction from topological corner states. (d) Energy spectrum of the supercell structure, exhibiting two in-gap TPS. (e) Charge density distributions of TPS 1 and 2 as indicated in (d).

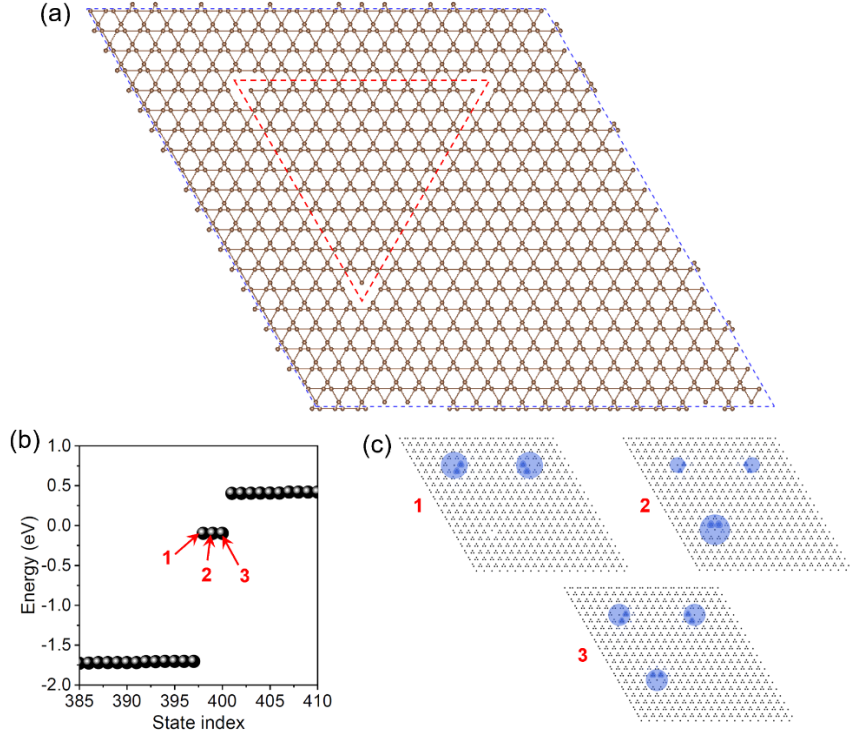

FIG. S13. (a) Supercell structure of the breathing Kagome lattice to construct the point-defects, which are effectively the 60° corners of the triangular flake highlighted by the red dashed lines. Here the periodic boundary conditions are adopted along the *a* and *b* directions to avoid the distraction from topological corner states. (b) Energy spectrum of the supercell structure, exhibiting three in-gap TPS. (c) Charge density distributions of TPS 1, 2, 3 as indicated in (b).

Puckered honeycomb lattice has been predicted as another platform to realize 2D HOTIs [14,15]. Here we choose phosphorene as an example of this family of materials to illustrate the topological corner states and TPS. Fig. S14(a) shows the structure of phosphorene, which is a distorted puckered honeycomb lattice. The energy band structure of phosphorene can be reproduced by a TB model with four orbitals ($s, p_x, p_y, p_z$) per site, whose Hamiltonian is written as $H = \sum_{i\alpha} \varepsilon_\alpha c_{i\alpha}^\dagger c_{i\alpha} + \sum_{\langle i\alpha, j\beta \rangle} t_{i\alpha,j\beta} c_{i\alpha}^\dagger c_{j\beta}$. Here $i, j$ and $\alpha, \beta$ are site and orbital indices respectively. $\varepsilon_\alpha$ in the first term is the onsite energy of orbital $\alpha$. The second term contains the NN hoppings which are parameterized following the Slater-Koster scheme. By using previous reported parameters of $\varepsilon_s = -17.10$, $\varepsilon_p = -8.33$, $t_{ss\sigma} = -1.40$, $t_{sp\sigma} = 1.84$, $t_{pp\sigma} = 3.24$, $t_{pp\pi} = -0.81$ eV [17], we calculate the bulk energy band structure as shown in Fig. S14(b). Then with this TB model, we calculate the energy spectrum of a finite rectangular flake and show the result in Fig. S14(c). The emerged in-gap states are topological corner states located at the top left and bottom right corners of the rectangular flake as shown in Fig. S14(d), consistent with previous studies [14].

Connecting the top left and bottom right corners, one can find the 1D nanostructure of phosphorene with first-order nontrivial topology. By cutting the 1D nanostructure into a 1D finite nanoribbon and solving the corresponding TB Hamiltonian, we obtain the energy spectrum of the finite nanoribbon as shown in Fig. S15(a), which shows the emergence of two states inside the bulk band gap of phosphorene. The charge density distribution of the two states (Fig. S15(b)) further confirms that they are topological end states of the nanoribbon. To generate the TPS inside the pristine bulk of phosphorene, we construct the point-defects in a supercell structure as shown in Fig. S15(c), which effectively realizes the end terminations of the first-order topologically nontrivial nanoribbon. The energy spectrum of this supercell structure is plotted in Fig. S15(d), which exhibits two in-gap TPS located at the point-defects, as displayed in Fig. S15(e), demonstrating the existence of TPS in HOTIs realized by this lattice model.

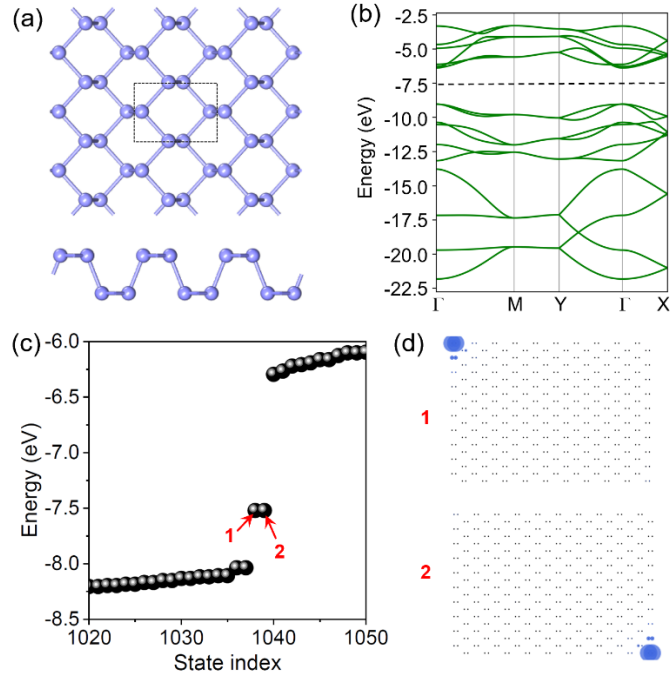

FIG. S14. (a) Structure of phosphorene as a representative for the puckered honeycomb lattice. The unit cell is represented by the dashed rectangle. (b) Bulk energy band structure of phosphorene by TB calculations. The dashed line indicates the Fermi energy. (c) Energy spectrum of the finite rectangular flake of phosphorene, exhibiting two in-gap topological corner states. (d) Charge density distributions of topological corner states 1, 2 as indicated in (c).

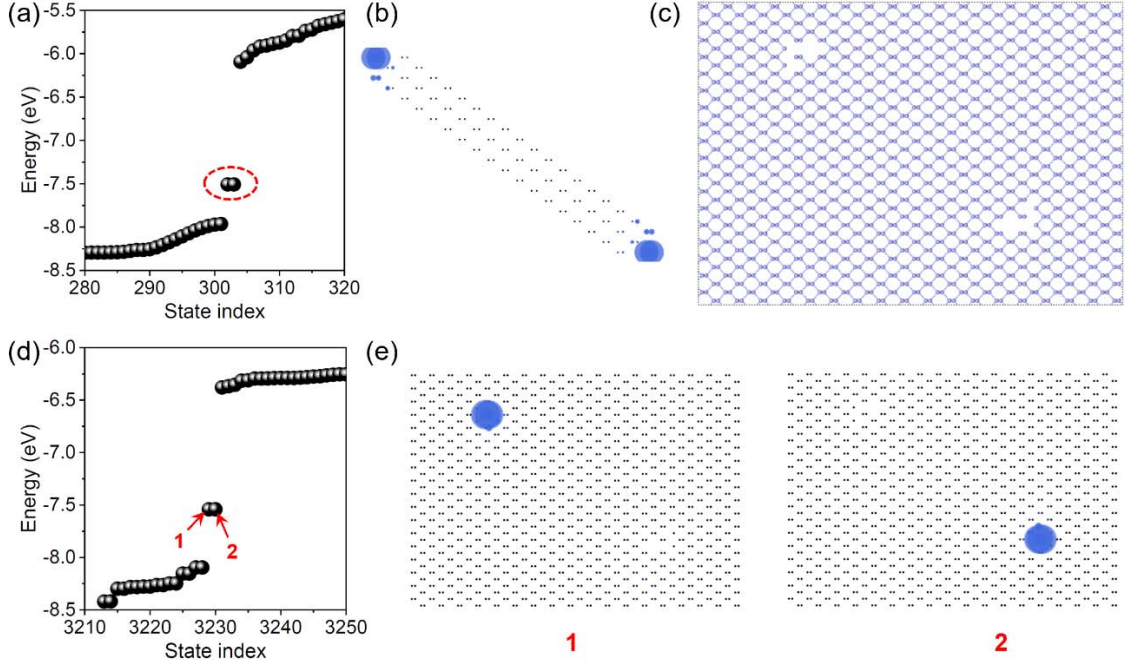

FIG. S15. (a) Energy spectrum of the finite 1D nanoribbon of phosphorene, showing the emergence of two topological end states inside the bulk band gap. (b) Charge density distribution of the topological end states. (c) Supercell structure of phosphorene containing two point-defects, which are effectively the ends of the finite nanoribbon. Here the periodic boundary conditions are adopted along the *a* and *b* directions to avoid the distraction from topological corner states. (d) Energy spectrum of the supercell structure, exhibiting two in-gap TPS. (e) Charge density distributions of TPS 1 and 2 as indicated in (d).

We also explore TPS in the HOTI realized by the Kane-Mele model [18,19] with an in-plane Zeeman field [16]. The honeycomb lattice of the Kane-Mele model is illustrated in Fig. S16(a). The single-orbital TB Hamiltonian reads

$$H = t\sum_{\langle ij \rangle} c_i^\dagger c_j + i\lambda_{SO} \sum_{\langle\langle ij \rangle\rangle} c_i^\dagger \hat{\boldsymbol{e}}_{ij} \cdot \boldsymbol{s} c_j + \sum_i c_i^\dagger \boldsymbol{B} \cdot \boldsymbol{s} c_i,$$

where the first and second terms correspond to the NN hopping and second NN spin-orbit coupling (SOC), respectively, and the third term is used to include the Zeeman field $\boldsymbol{B}$. $\boldsymbol{s}$ denotes the vector of spin Pauli matrices and $\hat{\boldsymbol{e}}_{ij}$ is the unit vector defined by $\hat{\boldsymbol{e}}_{ij} = (\vec{d}_{ij}^{\,1} \times \vec{d}_{ij}^{\,2})/|\vec{d}_{ij}^{\,1} \times \vec{d}_{ij}^{\,2}|$, where $\vec{d}_{ij}^{\,1}$ and $\vec{d}_{ij}^{\,2}$ are the two NN bond vectors connecting the second NN orbitals $j$ and $i$. Here we set $t = 1$ and $\lambda_{SO} = 0.1$ eV. It has been demonstrated that the application of a $y$-direction Zeeman field, i.e., $\boldsymbol{B} = (0, \lambda_y, 0)$, can tune the system from a $Z_2$ topological insulator into a HOTI [16]. In Fig. S16(b), we show the calculated band structures of a zigzag nanoribbon without ($\lambda_y = 0$) and with ($\lambda_y = 0.2$ eV) the Zeeman field. One can see that, the topological edge states become gapped in the presence of the in-plane Zeeman field. As such, the energy spectrum of a rhombus flake exhibits two in-gap topological corner states located at the 120° corners of the flake, as displayed in Fig. S16(c) and (d), demonstrating the phase transition from a TI to a HOTI induced by the in-plane Zeeman field.

By connecting the 120° corners of the flake, one can obtain the first-order topologically nontrivial 1D nanostructure of the 2D HOTI. Fig. S17(a) shows the energy spectrum of the finite nanoribbon which is cut from the 2D rhombus flake. One can see that there emerge two in-gap states, whose charge density distribution is located at the ends of the finite nanoribbon (Fig. S17(b)), demonstrating that they are topological end states. To generate the TPS, we construct a rhombus flake containing two point-defects which are effectively the ends of the finite nanoribbon (Fig. S18(a)). However, the calculated energy spectrum only shows two in-gap topological corner states. The emerged TPS are found to merge into the bulk states, as indicated in Fig. S18(b) and (c). This result is caused by the special symmetry of the lattice, as explained below. When constructing the point-defects following the end terminations of the nanoribbon, we obtain the point-defects with geometries as illustrated in Fig. S18(d). Because of the lattice symmetry, the TPS always appear in pair at one point-defect. Since these two TPS are spatially very close to each other, they will overlap and interact with each other to open a large band gap as illustrated in Fig. S18(e), pushing these states to merge into the bulk states. Nevertheless, we demonstrate that by increasing the spatial distance between the two TPS of one point-defect, the interaction between these states will be suppressed and the energy levels of TPS will be tuned into the bulk band gap, as displayed in Fig. S19(a)-(c). Furthermore, by completely removing one of the two TPS for each point-defect, one can obtain the ideal TPS located inside the bulk band gap as verified in Fig. S19(d) and (e).

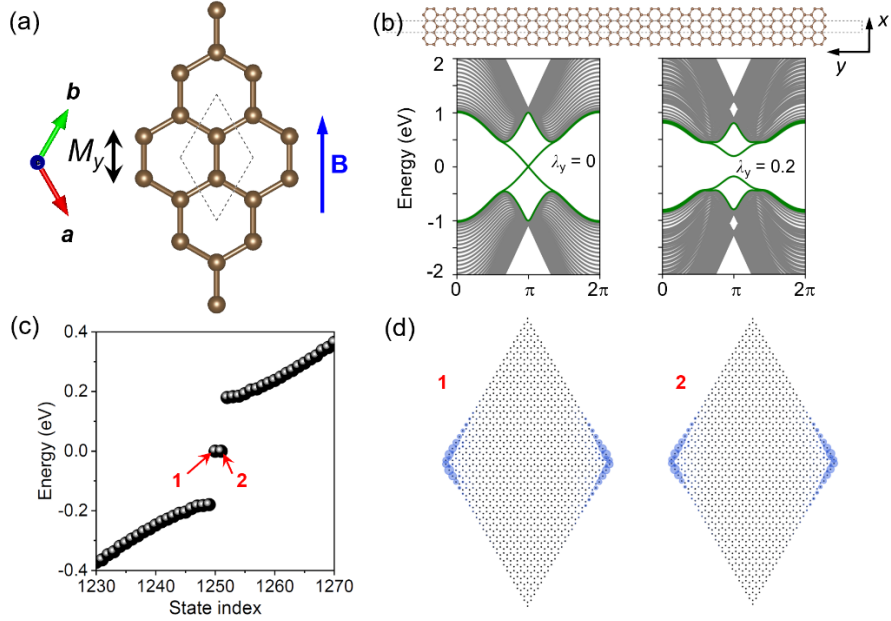

FIG. S16. (a) Structure of the Kane-Mele model in the honeycomb lattice. The unit cell is represented by the dashed rhombus. The in-plane Zeeman field **B** is applied along the **y** (armchair) direction, which preserves the mirror-refection symmetry $M_y$. (b) Structure of the zigzag nanoribbon and its energy band structures without and with the application of the in-plane Zeeman field. (c) Energy spectrum of the finite rhombus flake, exhibiting two in-gap topological corner states. (d) Charge density distributions of topological corner states 1, 2 as indicated in (c).

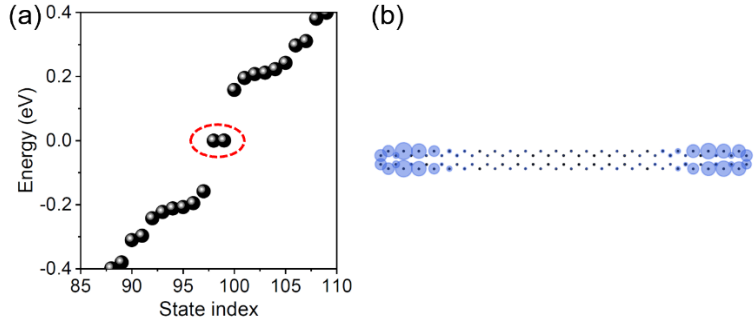

FIG. S17. (a) Energy spectrum of the finite nanoribbon of the Kane-Mele model in the presence of the in-plane Zeeman field, showing the emergence of two topological end states inside the bulk band gap. (b) Charge density distribution of the topological end states.

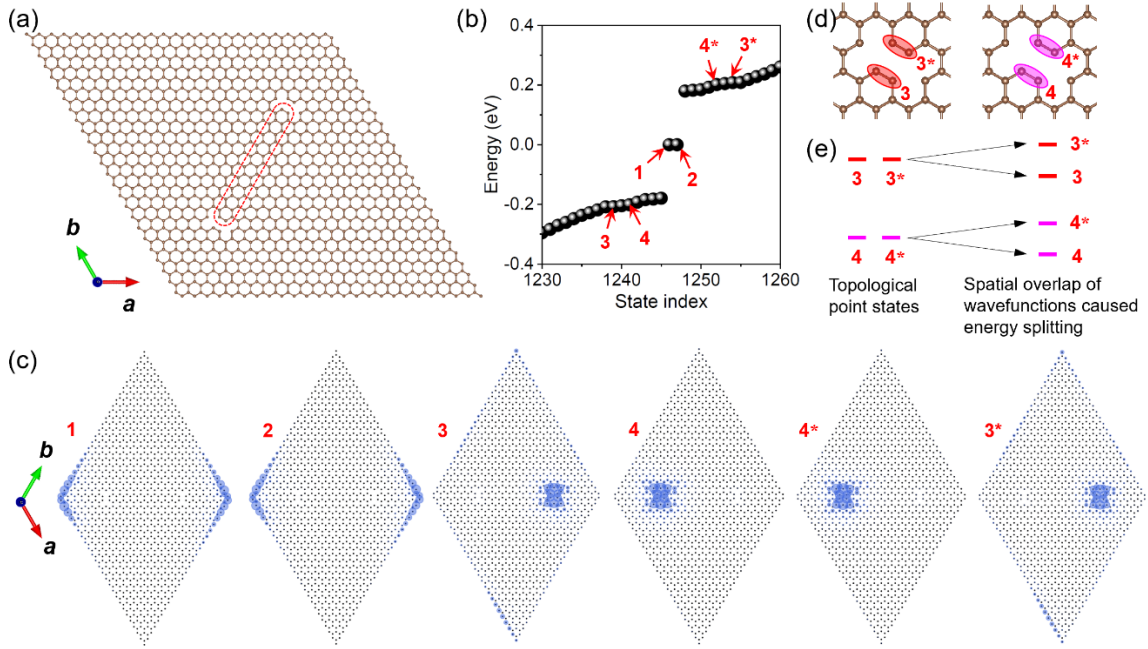

FIG. S18. (a) Supercell structure of the Kane-Mele model containing two point-defects, which are effectively the ends of the finite nanoribbon as highlighted by the red dashed lines. Here the open boundary conditions are adopted along all directions. (b) Energy spectrum of the finite supercell structure, exhibiting two in-gap topological corner states (1, 2) and four TPS (3, 4, 3* and 4*) hiding in the bulk states. (c) Charge density distributions of topological corner states and TPS as indicated in (b). (d) Schematic of spatial distributions of TPS before hybridization. States 3 and 3* are originated from the right point-defect while 4 and 4* are from the left point-defect. (e) Illustration of energy splitting of TPS caused by wavefunctions overlap and hybridization. Because of lattice symmetry, for each point-defect, there are two TPS, like 3 and 3* for the right point-defect. These two states are spatially very close to each other, therefore have significant wavefunctions overlap and hybridization, which causes the large energy splitting and pushes the TPS submerged into the bulk states.

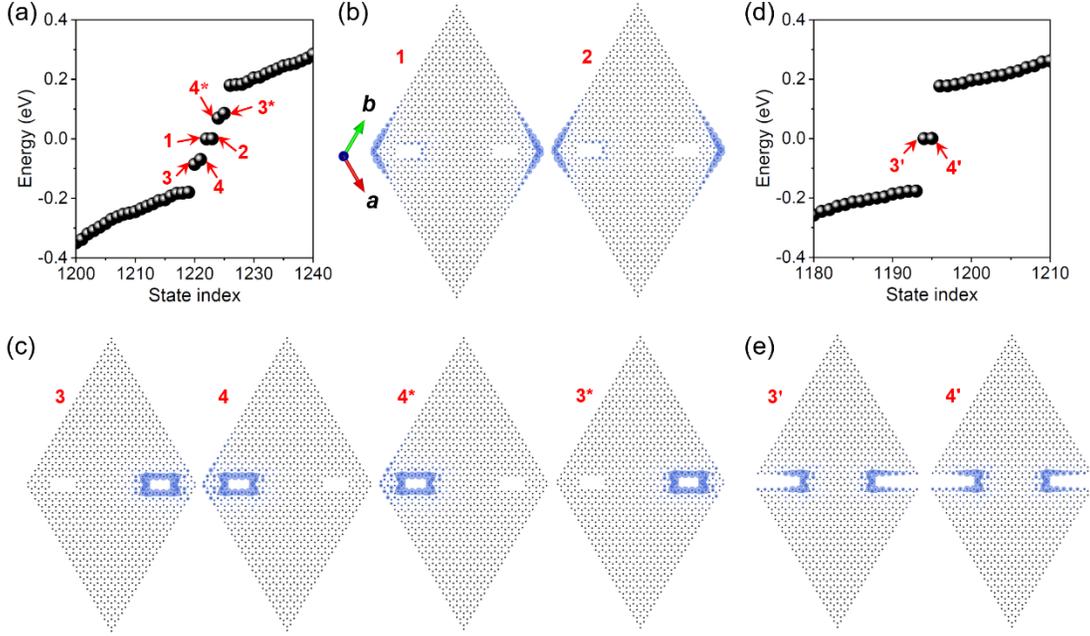

FIG. S19. (a) Energy spectrum of the finite supercell structure, where the spatial scale of each point-defect is increased to suppress the wavefunctions overlap of TPS from the same point-defect. The geometrical structure of the finite flake is as shown in (b). In this case, the energy splitting between TPS are reduced, and the four TPS are pushed back inside the bulk band gap. (b) and (c) are charge density distributions of the topological corner states and TPS as indicated in (a). (d) Energy spectrum of the finite supercell structure as shown in (e). In this case, since one of the two TPS for each point-defect and the 120° corners are all completely removed, there are only two ideal TPS emerged in the bulk band gap. (e) Charge density distributions of the TPS as indicated in (d).

## VIII. Illustration of the local region covered by the Zeeman field

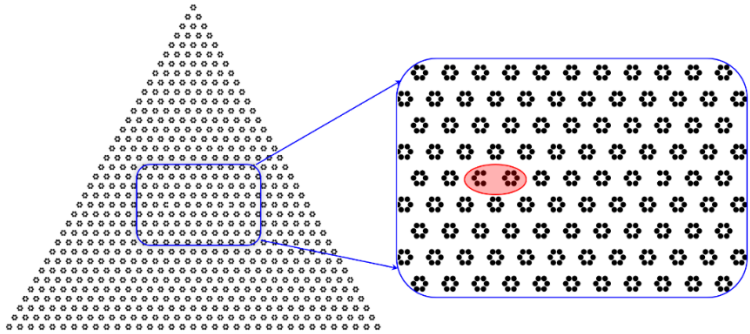

FIG. S20. Illustration of the local region where the Zeeman field is applied for Fig. 3 of the main text. The left panel is the geometrical structure of the finite triangular flake. The right plane is the zoomed-in plot near the point-defects. The red shadow indicates the region covered by the local Zeeman field.

# IX. Tight-binding model for γ-graphyne

The TB Hamiltonian for γ-graphyne based on a single-orbital effective lattice model (Fig. S21(a)) is written as $H = -t_{intra} \sum_{\langle ij \rangle_{intra}} c_i^\dagger c_j - t_{inter} \sum_{\langle ij \rangle_{inter}} c_i^\dagger c_j$. The first and second terms correspond to the intra- and inter-cell NN hoppings respectively. By fitting DFT band structure as shown in Fig. S21(b), we derive the TB parameters of $t_{intra}$ = 1.5 and $t_{inter}$ = -1.75 eV. With this TB model, we calculate the energy spectrum of a supercell structure containing two point-defects as shown in Fig. 4(a) of the main text and plot the result in Fig. S21(c). One can see that the TB energy spectrum is in good agreement with the DFT result. Moreover, the corresponding charge density distributions are also consistent between the TB and DFT calculations. These results validate the accuracy of the effective TB model. Based on this TB model, the spatial distribution of topological corner states and the first-order topologically nontrivial 1D nanostructure of γ-graphyne are displayed in Fig. S22.

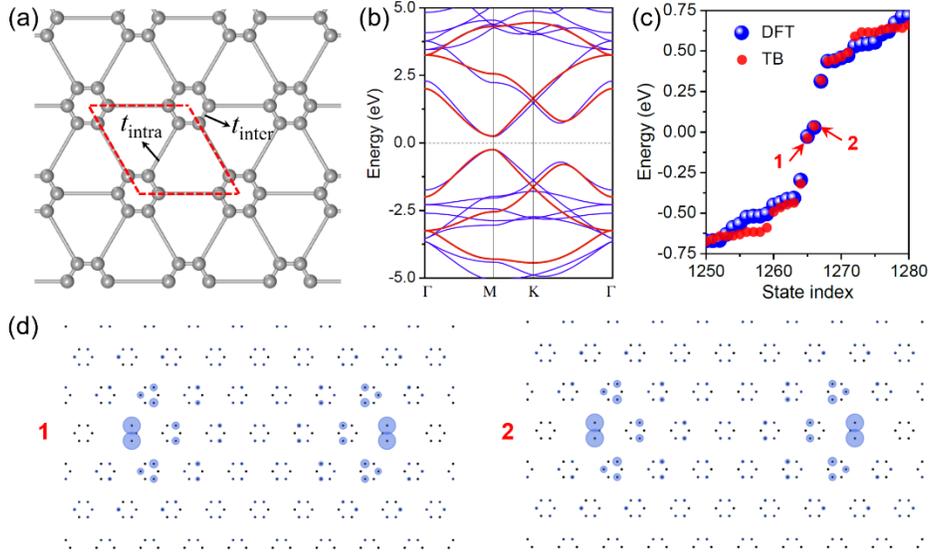

FIG. S21. (a) Simplified γ-graphyne lattice structure with hoppings of $t_{intra}$ and $t_{inter}$ for TB calculations. (b) Bulk energy band structure of γ-graphyne. Blue and red lines are results by DFT and TB calculations respectively. (c) Energy spectrum of the supercell structure of γ-graphyne with point-defects as shown in Fig. 4(a) of the main text. Blue and red dots are results by DFT and TB calculations, both showing the emergence of TPS. (d) Charge density distributions of TPS as indicated in (c). These results are obtained by TB calculations, which agree well with DFT results as shown in Fig. 4(c) of the main text.

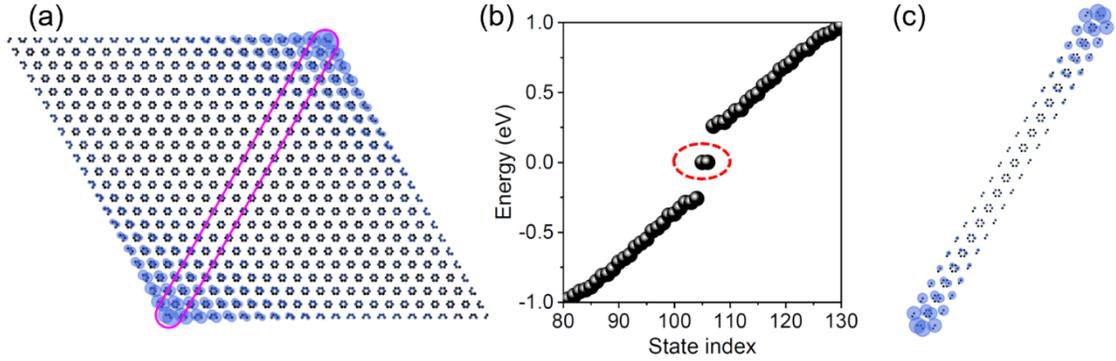

FIG. S22. (a) Spatial distribution of topological corner states of a rhombus flake of γ-graphyne. The purple line indicates the first-order topologically nontrivial 1D nanostructure of γ-graphyne. (b) Energy spectrum of the finite 1D nanostructure cut from the rhombus flake, exhibiting two topological end states whose spatial distribution is plotted in (c).

To investigate TPS in a large-scale sample of γ-graphyne, we turn to TB model analyses. The energy evolutions of TPS as a function of spatial separation L calculated using the TB method are plotted in Fig. S23(a), showing clearly visible energy splitting for L ≤ 10 (corresponding to ~ 6.89 nm). Moreover, we modelled the effect of a local Zeeman field. Specifically, we propose to apply a local Zeeman field by using a magnetic tip [20] approaching one point-defect, while using another nonmagnetic tip to measure the dI/dV curve at the other point-defect, as illustrated in Fig. S23(b). The field strength can be enhanced by moving the magnetic tip closer to the defect [21]. To simulate this process, we choose two TPS with L = 8 (~ 5.51 nm), whose energy splitting is estimated to be ~20 meV by TB calculation. We note that this splitting is likely underestimated because the TB parameters are fit to the DFT bands which are known for underestimating band gap [3,4]. In Fig. S23(c) we plot the evolution of local DOS of the right point-defect when the left point-defect is approached by the magnetic tip, to mimic the dI/dV curves probed by the nonmagnetic tip. Clearly, the two peaks merge into one single peak around the Fermi energy with the increasing Zeeman field, which agrees well with our model predictions.

To validate the feasibility of applying a local Zeeman field, we consider the adsorption of a single magnetic atom V at the left point-defect and perform DFT calculations to check the response of TPS. To do so, we construct a larger γ-graphyne supercell containing two point-defects as shown in Fig. S24(a), where the spatial separation between the two point-defects is ~ 3.3 nm. Then, a V atom is introduced at the TPS as shown in Fig. S24(d). DFT calculations are performed for both setups, as shown in Fig. S24(b)-(c) and (e)-(f) respectively. One sees that the V atom splits the energy of the left TPS, and the energy splitting of the right TPS is tuned much smaller, consistent with our model study results. We note that DFT results unambiguously confirmed the Zeeman field induced by V atom on the left TPS, same as the local field used in our model, and a pronounced reduction of energy splitting of the right TPS, as predicted by our model.

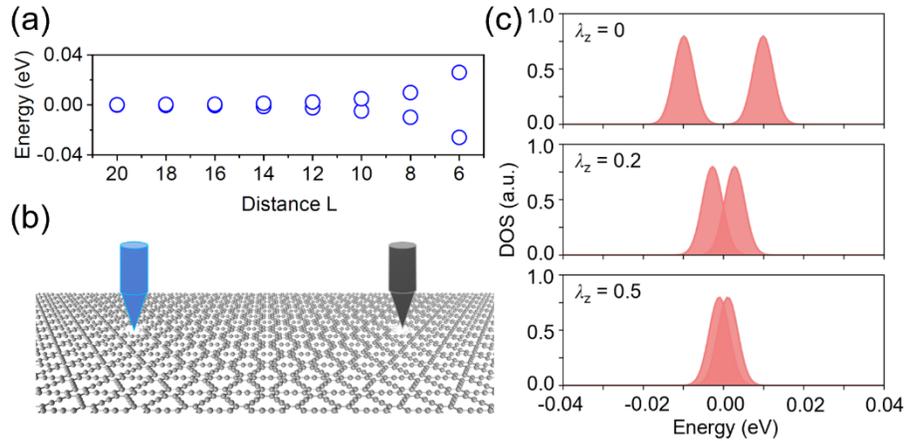

FIG. S23. (a) Energy evolution of TPS as a function of spatial separation L, in units of lattice constant of γ-graphyne. (b) Schematic experimental setup with two tips to probe the energy evolution of the TPS. The blue magnetic tip applies a local Zeeman field to the left point-defect, while the black nonmagnetic tip is used to measure the dI/dV curve of the right point-defect. (c) DOS of the TPS originated from the right point-defect when the strength of the local Zeeman field applied to the left point-defect increases. Here L = 8 (~ 5.51 nm).

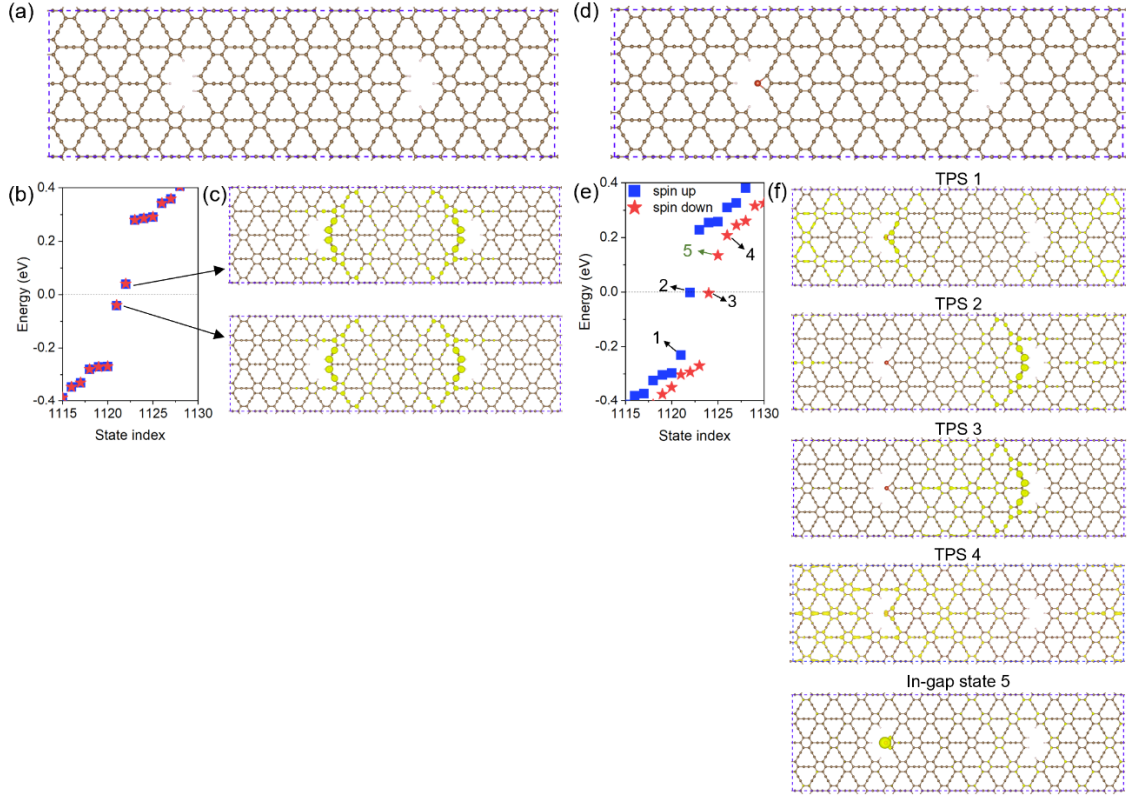

FIG. S24. (a) Supercell structure of γ-graphyne containing two point-defects. (b) Energy spectrum of the supercell, showing the emergence of TPS located inside the bulk band gap. (c) Charge density distributions of in-gap TPS (yellow). (d)-(f) Same as (a)-(c) but with the adsorption of a V atom at the left point-defect. Compared with the result before V-adsorption, the local Zeeman field induced by the magnetic atom is sufficient to split the energy of the left TPS, and the energy splitting of the right TPS is reduced, consistent with the results of model study.

## X. Selective hydrogenation of γ-graphyne to obtain topological point states

Besides creating vacancies, it is also possible to obtain TPS by adsorption of hydrogen or halogen atoms on selected atoms of a 2D sample, to filter out the valence orbitals on these atoms and remove their bonding with neighboring atoms, acting effectively like a vacancy [22,23], as demonstrated by our DFT calculations in Fig. S25.

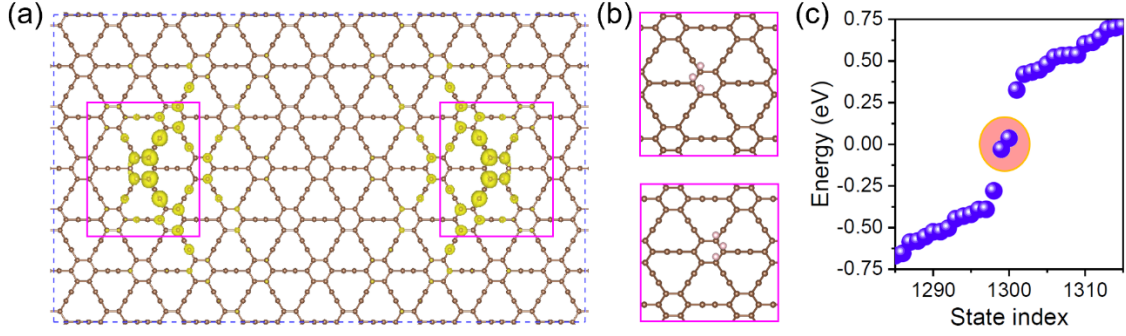

FIG. S25. (a) Supercell structure of γ-graphyne with effective "vacancies" by adsorption of hydrogen atoms (pink) on selected carbon atoms (bronze). (b) Zoomed-in structures near the effective "vacancies". The top and bottom panel are for the left and right "vacancy" highlighted by the purple rectangles in (a), respectively. (c) Energy spectrum at the Brillouin-zone-center Γ point of the supercell structure, exhibiting in-gap TPS marked by the red ellipse. Charge density distributions of in-gap TPS are plotted in (a) (yellow).

## XI. Tunable number and location of topological point states

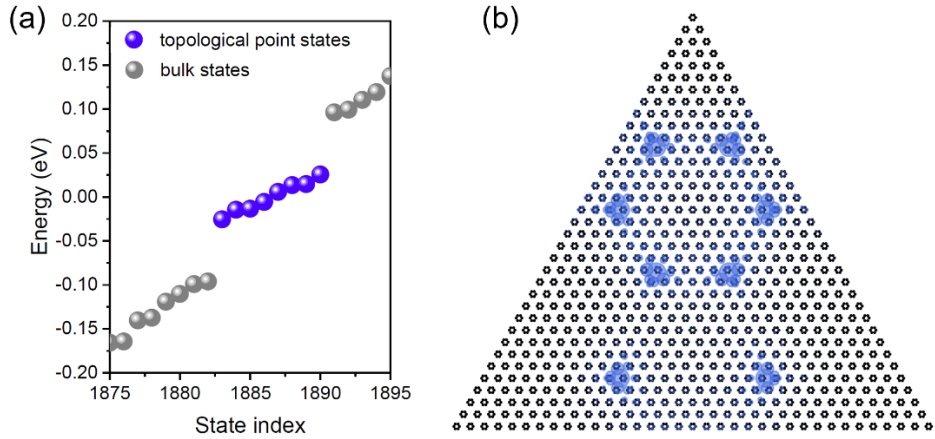

FIG. S26. (a) Energy spectrum of the triangular-shaped finite flake of the Kekulé lattice with $t_{intra}$ = 1 and $t_{inter}$ = 1.2 eV. In this flake, multiple point-defects are constructed, effectively realizing a total of eight topological corners of hexagonal and rhombus flakes in one single sample. With these point-defects, the energy spectrum exhibits eight TPS, whose charge density distributions can be visualized in (b), demonstrating that the proposed TPS are much easier to be created and tuned in terms of both number and position separately compared to topological corner states which can only at the corners and be changed by changing the whole sample shape.

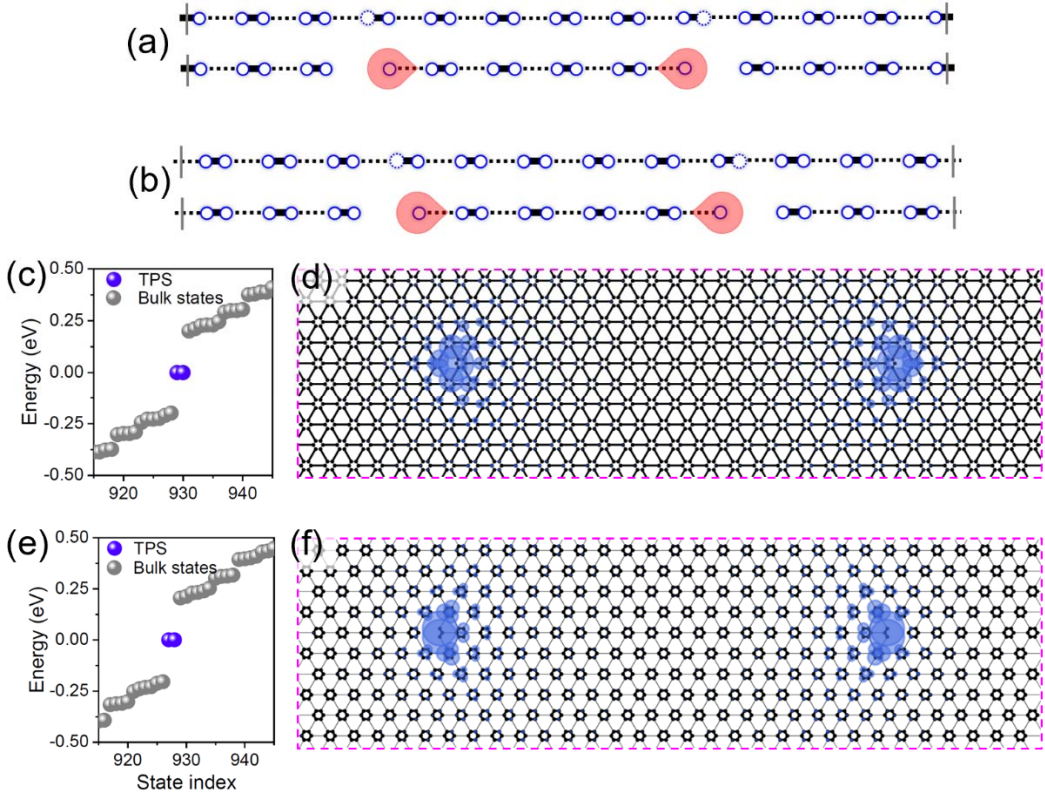

FIG. S27. (a) and (b) illustrate the emergence of TPS in both topological nontrivial and trivial phases of 1D SSH chain respectively. Solid and dashed lines represent strong and weak bonds. By shifting the position of vacancies, the middle section is always topological nontrivial and gives rise to topological "end" states. (c) and (e) Energy spectra of the periodic rectangle supercell with two vacancies of the topological nontrivial ($t_{intra}$ = 1 and $t_{inter}$ = 1.2 eV) and trivial ($t_{intra}$ = 1.2 and $t_{inter}$ = 1 eV) Kekulé lattice respectively, both showing the emergence of TPS. Corresponding charge density distributions of TPS are plotted in (d) and (f) in blue. For simplicity, periodic boundary condition is adopted to avoid the interference of dangling-bond states originated from outer boundaries.

**References**

[1] G. Kresse and J. Furthmüller, Phys. Rev. B **54**, 11169 (1996).
[2] P. E. Blöchl, Phys. Rev. B **50**, 17953 (1994).
[3] J. P. Perdew, K. Burke, and M. Ernzerhof, Phys. Rev. Lett. **77**, 3865 (1996).
[4] J. P. Perdew, K. Burke, and M. Ernzerhof, Phys. Rev. Lett. **78**, 1396 (1997).
[5] H. J. Monkhorst and J. D. Pack, Phys. Rev. B **13**, 5188 (1976).
[6] L.-H. Wu and X. Hu, Sci. Rep. **6**, 24347 (2016).
[7] Y. Liu, C.-S. Lian, Y. Li, Y. Xu, and W. Duan, Phys. Rev. Lett. **119**, 255901 (2017).
[8] B. Liu, G. Zhao, Z. Liu, and Z. F. Wang, Nano Lett. **19**, 6492 (2019).
[9] F. Zangeneh-Nejad and R. Fleury, Phys. Rev. Lett. **123**, 053902 (2019).
[10] T. Mizoguchi, H. Araki, and Y. Hatsugai, J. Phys. Soc. Jpn. **88**, 104703 (2019).

[11] Y. Zhou and R. Wu, Phys. Rev. B **107**, 035412 (2023).
[12] S.-Q. Shen, *Topological insulators* (Springer, 2012), Vol. 174.
[13] M. Ezawa, Phys. Rev. Lett. **120**, 026801 (2018).
[14] S. Qian, C.-C. Liu, and Y. Yao, Phys. Rev. B **104**, 245427 (2021).
[15] M. Pan, D. Li, J. Fan, and H. Huang, Npj Comput. Mater. **8**, 1 (2022).
[16] Y. Ren, Z. Qiao, and Q. Niu, Phys. Rev. Lett. **124**, 166804 (2020).
[17] W. A. Harrison, *Electronic Structure and the Properties of Solids: The Physics of the Chemical Bond* (Dover Publications, Newburyport, 2012).
[18] C. L. Kane and E. J. Mele, Phys. Rev. Lett. **95**, 146802 (2005).
[19] C. L. Kane and E. J. Mele, Phys. Rev. Lett. **95**, 226801 (2005).
[20] R. Wiesendanger, D. Bürgler, G. Tarrach, T. Schaub, U. Hartmann, H. J. Güntherodt, I. V. Shvets, and J. M. D. Coey, Appl. Phys. A **53**, 349 (1991).
[21] Z. Zhang, X. Ni, H. Huang, L. Hu, and F. Liu, Phys. Rev. B **99**, 115441 (2019).
[22] M. Zhou, W. Ming, Z. Liu, Z. Wang, P. Li, and F. Liu, Proc. Natl. Acad. Sci. U.S.A. **111**, 14378 (2014).
[23] H. Zhang, Y. Wang, W. Yang, J. Zhang, X. Xu, and F. Liu, Nano Lett. **21**, 5828 (2021).